\documentclass[preprint,prd,showpacs]{revtex4}
\usepackage[draft=false,letterpaper=true,colorlinks=true,urlcolor=black,linkcolor=black,citecolor=black,bookmarksopen=false,bookmarksnumbered=true,breaklinks=true,pdfauthor={Antonio Mondragon},pdfcreator={Antonio Mondragon},pdftitle={A Keplerian Limit to Static Spherical Spacetimes in Curvature Coordinates},pdfkeywords={Schwarzschild,Reissner,Nordstrom,de Sitter,Kottler,precession,Keplerian,relativity,blackhole,spacetime,Einstein,cosmological constant,charge,anti-de Sitter}]{hyperref}
\usepackage[all]{hypcap}
\usepackage{amsmath,amssymb}
\usepackage{verbatim}
\usepackage{graphicx,color}
\def\epm{\epsilon_{\!M}^{}}
\def\epq{\epsilon_{\hspace*{-0.2mm}Q}^{}}
\def\epL{\epsilon_{\!\Lambda}^{}}
\def\epmsquared{\epsilon_{\!M}^2}
\def\epqsquared{\epsilon_{\hspace*{-0.2mm}Q}^2}
\def\epqcubed{\epsilon_{\hspace*{-0.2mm}Q}^3}
\def\epmcubed{\epsilon_{\!M}^3}
\def\epmfourth{\epsilon_{\!M}^4}
\def\mpe{\epsilon_{\!M}^{-1}}

\def\rq{r_{\!Q}^{}}
\def\rqsquared{r_{\!Q}^2}
\def\rL{r_{\!\Lambda}^{}}
\def\rLsquared{r_{\!\Lambda}^2}
\def\rm{r_{\!M}^{}}
\def\rmsquared{r_{\!M}^2}
\begin{document}
\title{A Keplerian Limit to Static Spherical Spacetimes in Curvature Coordinates}
\date{\today}
\author{Tyler J. Lemmon}
\author{Antonio R. Mondragon}
\email{antonio.mondragon@coloradocollege.edu}
\affiliation{Physics Department, Colorado College, Colorado Springs, Colorado 80903}
\begin{abstract}
The problem of a test body in the Schwarzschild geometry is investigated in a Keplerian limit.  Beginning with the Schwarzschild metric, a solution to the limited case of approximately elliptical (Keplerian) motion is derived in terms of trigonometric functions.  This solution is similar in form to that derived from Newtonian mechanics, and includes first-order corrections describing three effects due to general relativity:  precession; reduced radial coordinate; and increased eccentricity.  The quantitative prediction of increased eccentricity may provide an additional observational test of general relativity.  By analogy with Keplerian orbits, approximate orbital energy parameters are defined in terms of a relativistic eccentricity, providing first-order corrections to Newtonian energies for elliptical orbits.  The first-order relativistic equation of orbit is demonstrated to be a limiting case of a very accurate self-consistent solution.  This self-consistent solution is supported by exact numerical solutions to the Schwarzschild geometry, displaying remarkable agreement.  A more detailed energy parameterization is investigated using the relativistic eccentricity together with the apsides derived from the relativistic effective potential in support of the approximate energy parameters defined using only first-order corrections.  The methods and approximations describing this Keplerian limit are applied to more general static spherically-symmetric geometries.  Specifically, equations of orbit and energy parameters are also derived in this Keplerian limit for the Reissner-Nordstr\"{o}m and Schwarzschild-de\,Sitter metrics.
\end{abstract}
\maketitle

%%%%%%%%%%%%%%%%%%%%%%%%%
\section{Introduction}\label{Sec_Intro}
%%%%%%%%%%%%%%%%%%%%%%%%%

The problem of a test body in the Schwarzschild geometry is of fundamental importance in understanding orbital characteristics due to general relativity.   Detailed analyses and approximate analytical solutions of this geometry exist \cite{MTW,weinberg,rindler,PleKra,OR,dinverno,carroll1,ovanesyan,schild,BroCle,hartle,wang,stump,nobili,magnan,deliseo,brillgoel,dean}, emphasizing a first order approximate rate of precession of perihelia of Mercury and the unstable circular orbit and relativistic capture representative of extreme astrophysical environments.  The general problem has been solved in terms of Weierstra\ss\ elliptic functions by Hagihara \cite{hagihara}, and more recently by Kraniotis and Whitehouse \cite{KW1}.  Ashby \cite{ashby} has solved the limited case of approximately Keplerian motion in terms of Jacobian elliptic functions, ``...which are closer in spirit to trigonometric functions, with which most are familiar.''  In this paper, beginning with the Schwarzschild geometry, a solution to the limited case of approximately Keplerian motion is derived in terms of trigonometric functions.  This approximate solution lends itself to easy comparison with the familiar Keplerian orbits (ellipses) of Newtonian mechanics and is consistent with both the reduced radius of circular orbit, commonly derived from the Schwarzschild effective potential, and the observed rate of precession of perihelia of the inner planets, commonly derived perturbatively.  Additional insights regarding the sizes and shapes of bound relativistic orbits are provided, including a relativistic correction to eccentricity which may be subjected to observational tests.  By analogy with Keplerian orbits, a relativistic eccentricity is used to define a Schwarzschild energy parameter, providing corrections to Newtonian energies for Keplerian orbits.  The relativistic equation of orbit together with the energy parameterization comprise a simple model that is useful for a qualitative and quantitative understanding of the corrections to Keplerian orbits due to general relativity.  This model is easily extended to include more general static spherically-symmetric geometries.  Specifically, models are also derived for the Reissner-Nordstr\"{o}m and Schwarzschild-de\,Sitter metrics.

The methods and approximations describing a Keplerian limit are detailed in Sec.~\ref{Sec_S_KepLim}.  Beginning with the Schwarzschild geometry, an approximate equation of orbit is derived in terms of trigonometric functions.  When compared to that of a corresponding Keplerian orbit, this equation of orbit clearly displays three characteristics of relativistic orbits:  precession; reduced radial coordinate; and increased eccentricity.  These characteristics arise as first-order relativistic corrections to the familiar equation of orbit describing Keplerian orbits.  This solution is found to be valid for near-circular orbits requiring only small relativistic corrections.  Predictions of relativistic precession and reduced radius of circular orbit are in agreement with known results.  This provides confidence in a new relativistic correction to eccentricity.  (The possibility of observing relativistic corrections to eccentricity is discussed briefly in Sec.~\ref{SubSec_S_Char}.)  In addition, first-order relativistic corrections to Keplerian apsides are predicted, resulting in the conclusion that the overall size of a relativistic orbit is smaller than a corresponding Keplerian orbit.  By analogy with Newtonian mechanics, Schwarzschild energy parameters are defined using the virial theorem for circular orbits, and using a relativistic eccentricity for noncircular orbits, resulting in first-order relativistic corrections to Newtonian energies for Keplerian orbits.  This simple model is substantiated by a more detailed investigation in Sec.~\ref{Sec_S_self_cons}.

Again beginning with the Schwarzschild geometry, the derivation of a very accurate self-consistent relativistic equation of orbit is detailed in Sec.~\ref{Sec_S_self_cons}.  This self-consistent equation of orbit is identical in form to the more approximate equation of orbit derived in Sec.~\ref{Sec_S_KepLim} and predicts the same orbital characteristics.  However, corrections to Keplerian orbits due to general relativity are described more accurately.  For example, the predicted radius of circular orbit is identical to that derived by minimizing the relativistic effective potential.  This solution is accurate in predicting long-term behavior of Schwarzschild orbits for a large parameter space, as is demonstrated by comparisons with exact numerical solutions.  The more approximate equation of orbit derived in Sec.~\ref{Sec_S_KepLim} is shown to be a limiting case of this self-consistent solution.  A more detailed energy parameterization is investigated for comparison with the simple parameterization of Sec.~\ref{Sec_S_KepLim}.  This parameterization is constructed using the relativistic eccentricity derived from the self-consistent equation of orbit together with the relativistic apsides derived from the Schwarzschild effective potential.  The energy parameters of Sec.~\ref{Sec_S_KepLim} approximate well the results from this more detailed parameterization, lending value to the simpler approach and more approximate results.

The methods and approximations describing a Keplerian limit to the Schwarzschild geometry are applied to a more general class of static spherically-symmetric geometries in curvature coordinates \cite{PleKra,OR,dinverno,carroll2,SH2,GaoZha}.  Specifically, the path of a small test mass in a static spherical spacetime is taken to be described by the metric
\begin{gather}
\mathrm{d}s^2 \!= \mathrm{e}^{2\nu(r)} c^2 \mathrm{d}  t^2 - \mathrm{e}^{-2\mu(r)} \mathrm{d}r^2 - r^2 \mathrm{d}\varOmega^2, \label{eq_metric_general}
\end{gather}
where $\mathrm{d}\varOmega^2 = \mathrm{d}\theta^2 + \sin^2{\!\theta}\,\mathrm{d}\varphi^2$, and
\begin{gather}
\mathrm{e}^{2\nu(r)} = \mathrm{e}^{2\mu(r)} = 1 - \frac{2M}{r} + \frac{Q^2}{r^2} - \frac{1}{3}\Lambda r^2 > 0.
\end{gather}
The Reissner-Nordstr\"{o}m $(\Lambda = 0)$ geometry is considered in Sec.~\ref{Sec_RN_KepLim}, followed by the Schwarzschild-de\,Sitter $(Q = 0,\Lambda>0)$ geometry in Sec.~\ref{Sec_SdS_KepLim}.  These geometries have been shown to meet all of the conditions for physical acceptability \cite{delgaty,semiz,bronnikov,maartens,skmhh}.  In each of these cases the relativistic correction due to matter is considered to be the dominant contribution.  The resulting orbital equations are of the same form as that derived for the Schwarzschild geometry, including additional corrections for each geometry.  For examples: there is an additional contribution to relativistic precession due to charge opposite in direction to the contribution due to matter; there is also an additional contribution to relativistic precession due to $\Lambda$, but in the same direction as the contribution due to matter.  Numerical studies are provided only for the Schwarzschild geometry in order to establish the self-consistent approach.  A self-consistent treatment of the Schwarzschild-de\,Sitter geometry is intractable and is not pursued.  A concise summary of the results for the three geometries is given in Sec.~\ref{Sec_Summary}.

%%%%%%%%%%%%%%%%%%%%%%%%%
\section{Schwarzschild Orbits in Keplerian Limit}\label{Sec_S_KepLim}
%%%%%%%%%%%%%%%%%%%%%%%%%

The path of a small test mass near a spherically-symmetric central mass $M$ is uniquely described by the Schwarzschild geometry \cite{MTW,weinberg,rindler,PleKra,OR,dinverno,carroll1,ovanesyan,schild,BroCle,hartle,wang,stump,schwarzschild,droste}.
\begin{gather}
\mathrm{d}s^2 \!= ( 1 - \rm/r ) c^2\mathrm{d}t^2 - \frac{\mathrm{d}r^2}{1 - \rm/r} - r^2 \mathrm{d}\varOmega^2,
\end{gather}
where $\mathrm{d}\varOmega^2=\mathrm{d}\theta^2 + \sin^2{\!\theta}\,\mathrm{d}\varphi^2$.  The singularity at the Schwarzschild radius $\rm \equiv 2GM/c^2$ is irrelevant in the present context in which a solution far from the central mass is sought.  Consider orbits in the plane defined by $\theta=\pi/2$.  Parameterize timelike geodesics with $\mathrm{d}s^2=c^2\mathrm{d}\tau^2$, where $\tau$ is the proper time along the path of a test particle.  Then, with $\dot{x}^\mu\equiv\mathrm{d}x^\mu/\mathrm{d}\tau$, the equations of motion may be expressed as
\begin{align}
\ell &\equiv r^2\dot\varphi,\\
k &\equiv (1 - \rm/r)\,\dot{t},\\
\tfrac{1}{2}(k^2-1)c^2 &=\tfrac{1}{2}\dot{r}^2 + \tilde{V}_\text{eff},\label{eq_S_eom}
\end{align}
where a Schwarzschild effective potential is defined as
\begin{gather}
r^2_\text{c}\ell^{-2}\tilde{V}_\text{eff} \equiv - \frac{r_\text{c}}{r} + \frac{1}{2}\frac{r^2_\text{c}}{r^2} - \epm \frac{r^3_\text{c}}{r^3}.  \label{eq_S_eff_pot}
\end{gather}
The parameter $r_\text{c}\equiv\ell^2/GM$ is the radius of a circular orbit for a nonrelativistic particle with the same angular momentum $\ell$, and the mass-related relativistic correction parameter is defined as
\begin{gather}
\epm\equiv\biggl( \!\frac{GM}{\ell c} \biggr)^{\!2} = \frac{1}{2} \frac{\rm}{r_\text{c}}. \label{eq_matter}
\end{gather}
The Newtonian effective potential is recovered in the limit $\epm \rightarrow 0$,
\begin{gather}
r^2_\text{c}\ell^{-2}V_\text{eff} \equiv - \frac{r_\text{c}}{r} + \frac{1}{2}\frac{r^2_\text{c}}{r^2}.  \label{eq_N_eff_pot}
\end{gather}
Time is eliminated from Eq.~(\ref{eq_S_eom}) using \hbox{$\dot{r}=-\ell\mathrm{d}u/\mathrm{d}\varphi$,} where $u \equiv 1/r$.  An equation for the trajectory of a test particle is then obtained by differentiating once more with respect to $\varphi$,
\begin{gather}
\frac{\mathrm{d}^2}{\mathrm{d} \varphi^2} \frac{r_\text{c}}{r} + \frac{r_\text{c}}{r} = 1 + 3\epm \Big( \frac{r_\text{c}}{r} \Big)^{\!2}. \label{eq_S_eoo_DE}
\end{gather}
The conic-sections of Newtonian mechanics are recovered by setting $\epm = 0$ in Eq.~(\ref{eq_S_eoo_DE}),
\begin{gather}
\frac{\mathrm{d}^2}{\mathrm{d} \varphi^2} \frac{r_\text{c}}{r} + \frac{r_\text{c}}{r} = 1 \quad\implies\quad \frac{r_\text{c}}{r} = 1 + e\cos{\varphi}, \label{eq_N_eoo}
\end{gather}
where $e$ is the eccentricity of the orbit and is taken to be positive or zero.

For approximately Keplerian orbits it is convenient to linearize the equation of motion (\ref{eq_S_eoo_DE}) by making the change of variable
\begin{gather}
\frac{r_\text{c}}{r} - 1 \equiv \frac{1}{\sigma} \ll 1,\label{eq_linearize}
\end{gather}
so that the last term on the right-hand-side of Eq.~(\ref{eq_S_eoo_DE}) may be approximated as
\begin{gather}
\Big( \frac{r_\text{c}}{r} \Big)^{\!2} \approx 1 + \frac{2}{\sigma}.
\end{gather}
Equation~(\ref{eq_S_eoo_DE}) may now be expressed as
\begin{gather}
\frac{1}{3\epm}\frac{\mathrm{d}^2}{\mathrm{d} \varphi^2} \frac{1}{\sigma} + \frac{1-6\epm}{3\epm}\frac{1}{\sigma} \approx 1.
\end{gather}
Defining $\sigma_\text{c}\equiv (1-6\epm)/3\epm$ and making the additional change of variable $\alpha\equiv (1-6\epm)^\frac{1}{2}\varphi$ results in the familiar form
\begin{gather}
\frac{\mathrm{d}^2}{\mathrm{d} {\alpha}^2} \frac{\sigma_\text{c}}{\sigma} + \frac{\sigma_\text{c}}{\sigma} \approx 1.
\end{gather}
The solution is described by Eq.~(\ref{eq_N_eoo}),
\begin{gather}
\frac{\sigma_\text{c}}{\sigma} \approx 1   + A\cos{\alpha},\label{eq_eoo_sigma}
\end{gather}
where $A$ is an arbitrary constant of integration.  In terms of the original coordinates, Eq.~(\ref{eq_eoo_sigma}) becomes
\begin{gather}
\frac{\tilde{r}_\text{c}}{r} \approx 1 + \tilde{e} \cos{\tilde{\kappa}\varphi}, \label{eq_S_eoo}
\end{gather}
where
\begin{align}
\tilde{r}_\text{c} &\equiv r_\text{c}\frac{1-6\epm}{1 - 3\epm}, \label{eq_S_coeff_r}\\
\tilde{e} &\equiv \frac{3\epm A}{1 - 3\epm}, \\
\tilde{\kappa} &\equiv (1 - 6\epm)^{\frac{1}{2}}. \label{eq_S_coeff_phi}
\end{align}
According to the correspondence principle, the solution must reduce to that of Newtonian mechanics in the limit $\epm\rightarrow 0$.  Therefore,
\begin{gather}
3\epm A\equiv e\label{eq_corr_princ}
\end{gather}
is identified as the eccentricity of Newtonian mechanics (\ref{eq_N_eoo}).  Then, to first order in $\epm$, Schwarzschild orbits in this Keplerian limit are described by Eq.~(\ref{eq_S_eoo}), where
\begin{align}
\tilde{r}_\text{c} &\approx r_\text{c} (1 - 3\epm), \label{eq_S_1st_coeff_r}\\
\tilde{e} &\approx e (1 + 3\epm), \label{eq_S_1st_coeff_e} \\
\tilde{\kappa} &\approx 1 - 3\epm.
\end{align}
Therefore, Schwarzschild orbits in this Keplerian limit may be expressed concisely as
\begin{gather}
\frac{r_\text{c}(1-3\epm)}{r} \approx 1 + e(1 + 3 \epm)\cos{(1 - 3 \epm)\varphi}. \label{eq_S_eoo_concise}
\end{gather}

A systematic verification may be carried out by substituting (\ref{eq_S_eoo_concise}) into (\ref{eq_S_eoo_DE}), keeping terms of orders $e$, $\epm$, and $e\epm$ only.  However, the justification for discarding the term nonlinear in eccentricity is the correspondence principle.  Arguments concerning which terms to discard based only on direct comparisons of relative magnitudes of higher-order and lower-order terms lead to contradictions.  Rather, the domain of validity is expressed by subjecting the solution (\ref{eq_S_eoo_concise}) to condition (\ref{eq_linearize}) for the smallest value of $r$.  Evaluating the equation of orbit (\ref{eq_S_eoo_concise}) at pericenter $(r = \tilde{r}_{-})$ results in
\begin{gather}
\frac{r_\text{c}}{\tilde{r}_{-}} = \frac{1 + e(1 + 3\epm)}{1 - 3\epm},
\end{gather}
so that, according to (\ref{eq_linearize}), the domain of validity is given by
\begin{gather}
e(1 + 3\epm) + 2(3\epm) \ll 1. \label{eq_S_1st_valid}
\end{gather}
Therefore, the relativistic eccentricity $\tilde{e} \equiv e(1+3\epm)\ll 1$; the Schwarzschild equation of orbit (\ref{eq_S_eoo_concise}) is limited to describing relativistic corrections to near-circular, elliptical orbits.  Also, the relativistic correction $6\epm\ll 1$;  the Schwarzschild equation of orbit (\ref{eq_S_eoo_concise}) is valid only for small relativistic corrections.

Although the more general solution, Eq.~(\ref{eq_S_eoo}) together with (\ref{eq_S_coeff_r})-(\ref{eq_S_coeff_phi}), contains higher-order terms in $\epm$, it is only consistent to first order.  A self-consistent solution that is consistent to all orders in $\epm$ and accurately predicts long-term orbital behavior is presented in Sec.~\ref{Sec_S_self_cons}.  However, the first-order relativistic equation of orbit (\ref{eq_S_eoo_concise}) is similar in form to that describing Keplerian orbits (\ref{eq_N_eoo}) and is useful for a qualitative and quantitative understanding of relativistic corrections to Keplerian orbits.  When compared to Keplerian orbits, Schwarzschild orbits clearly display three characteristics:  precession; reduced radial coordinate; and increased eccentricity.

%%%%%%%%%%%%%%%%%%%%%%%%%
\subsection{Characteristics of Schwarzschild Orbits}\label{SubSec_S_Char}
%%%%%%%%%%%%%%%%%%%%%%%%%

\begin{figure}
\centering
\includegraphics[width=\columnwidth]{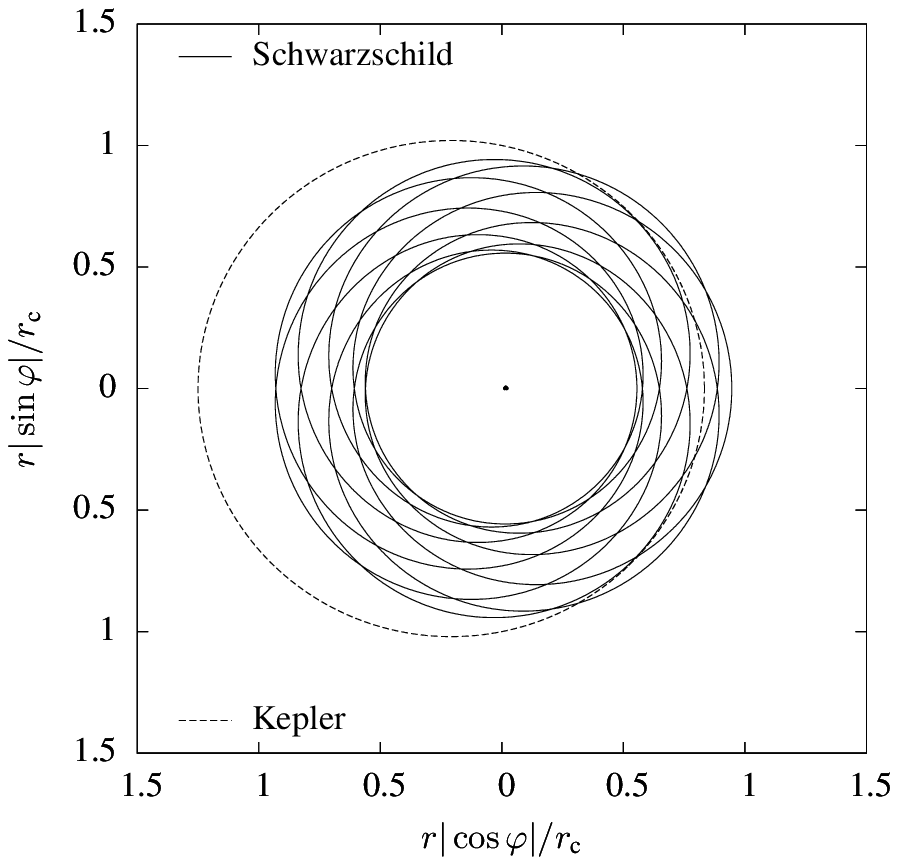}%
\caption{\label{fig_precess} Central-mass orbit derived from the Schwarzschild geometry (solid), Eq.~(\ref{eq_S_eoo_concise}), compared to a corresponding Keplerian orbit (dashed), Eq.~(\ref{eq_N_eoo}).  Precession is one orbital characteristic due to relativity, and is illustrated here for $0\leq\varphi\leq 20\pi$.  The eccentricity is chosen to be $e=0.2$ for both the Schwarzschild and Keplerian orbits.  This effect is exaggerated by the choice of relativistic correction parameter $(\epm=0.1)$ for purposes of illustration.  However, precession is present for smaller (non-zero), reasonably chosen values of $\epm$ as well.}
\end{figure}

The approximate equation of orbit (\ref{eq_S_eoo_concise}) predicts a shift in apside through an angle
\begin{align}
\Delta\varphi &\equiv 2\pi (\tilde{\kappa}^{-1} - 1) \label{eq_S_def_precess} \\
&\approx 3\epm(2\pi) \label{eq_S_1st_precess}
\end{align}
per revolution.  This first-order prediction is in agreement with existing perturbative calculations \cite{PleKra,carroll2,HobEfsLas}, as well as with other calculations \cite{nobili,magnan,deliseo,brillgoel,dean}, and is well-known to be in agreement with the observed precession of perihelia of the inner planets \cite{stump,ovanesyan,stewart,sigismondi}.  Precession due to relativity is illustrated in Fig.~\ref{fig_precess}.  The rate of precession is exaggerated by the choice of relativistic correction parameter $(\epm=0.1)$ for purposes of illustration.  However, relativistic orbits precess for smaller (non-zero), reasonably chosen values of $\epm$ as well;  Figure~\ref{fig_precess} correctly illustrates that relativistic orbits precess, a characteristic which is not present in Keplerian (central-mass) orbits.

\begin{figure}
\centering
\includegraphics[width=\columnwidth]{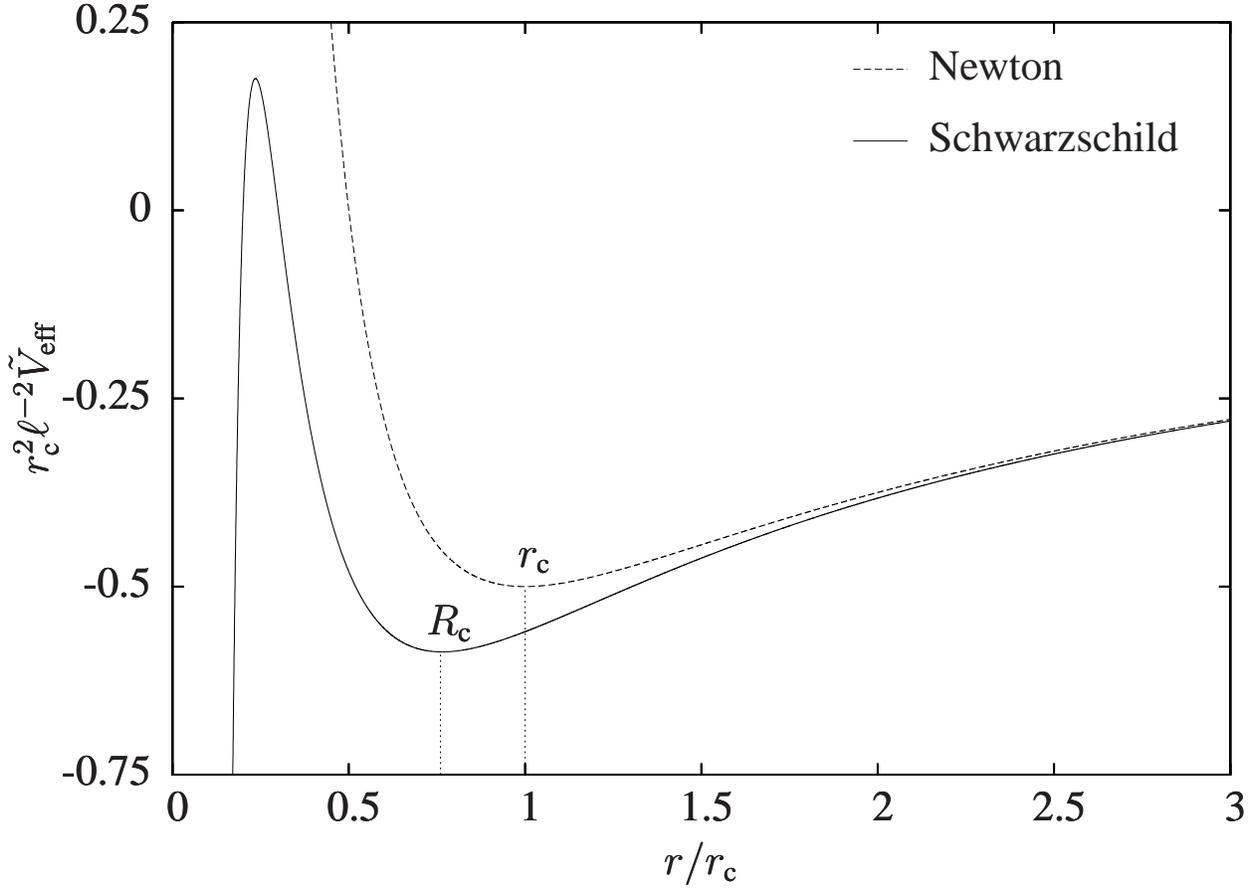}%
\caption{\label{fig_S_eff_pot} Effective potential (\ref{eq_S_eff_pot}) derived from the Schwarzschild geometry (solid) compared to that derived from Newtonian mechanics (dashed).  The vertical dotted lines identify the radius of circular orbit as predicted by Schwarzschild $R_\text{c}$, and Newton $r_\text{c}$.  The Schwarzschild geometry predicts a smaller radius of circular orbit as given by Eq.~(\ref{eq_S_rc_eff_pot}).  The value $\epm=0.06$ has been chosen for purposes of illustration.  This large value is inconsistent with the condition $\epm \ll 1/12$ and, therefore, with the approximation of Eq.~(\ref{eq_S_1st_r}).    However, a self-consistent solution derived in Sec.~\ref{Sec_S_self_cons} correctly predicts the radius of circular Schwarzschild orbit (\ref{eq_S_SC_coeff_r}) for this large relativistic correction parameter.}
\end{figure}

The approximate equation of orbit (\ref{eq_S_eoo_concise}) predicts a reduced radius of circular orbit,
\begin{gather}
\delta \tilde{r}_\text{c} \equiv \tilde{r}_\text{c} - r_\text{c} \approx - 3 \epm r_\text{c}. \label{eq_S_1st_r}
\end{gather}
This prediction is in agreement \cite{MTW,carroll1,wald} to first order with that obtained by minimizing the Schwarzschild effective potential (\ref{eq_S_eff_pot}),
\begin{gather}
R_\text{c} = \frac{1}{2} r_\text{c} + \frac{1}{2} r_\text{c} \sqrt{1 - 12 \epm}. \label{eq_S_rc_eff_pot}
\end{gather}
(A distinction is made here between the radius of circular orbit as determined from the Schwarzschild effective potential $R_\text{c}$ and that determined from the equation of orbit $\tilde{r}_\text{c}$.)  For $\epm\ll 1/12$, the radius of circular orbit is predicted to be reduced,
\begin{gather}
\delta R_\text{c} \equiv R_\text{c} - r_\text{c} \approx - 3 \epm r_\text{c}.
\end{gather}
The Schwarzschild effective potential (\ref{eq_S_eff_pot}) is compared to that derived from Newtonian mechanics (\ref{eq_N_eff_pot}) in Fig.~\ref{fig_S_eff_pot}.

The approximate equation of orbit (\ref{eq_S_eoo_concise}) provides the further insight that a non-circular Schwarzschild orbit is predicted to be smaller than a corresponding Keplerian orbit for a large range of eccentricities.  The relativistic apsides may be expressed approximately as
\begin{gather}
\tilde{r}_\pm/r_\pm \approx 1-3\epm\frac{1 \mp 2e}{1 \mp e},
\end{gather}
where $r_\pm \equiv r_\text{c}/(1 \mp e)$ denotes the Keplerian apocenter $(+)$ and pericenter $(-)$.  The relativistic apocenter is equal to the corresponding Keplerian apocenter for $e_{+}\approx1/2$.  For all nonzero values of $\epm$, the apocenter is reduced for $e<e_{+}$ and increased for $e>e_{+}$.  The relativistic pericenter is reduced for all nonzero values of $\epm$.  Therefore, for $e<e_{+}$, the radial coordinate is reduced;  the observed overall size of a relativistic orbit is smaller than a corresponding Keplerian orbit for $e<e_{+}$.  Using a perturbative treatment, Nobili and Roxburgh identified a correction of the same order of magnitude \cite{nobili}:
\begin{quote}
``There is a constant part, causing a variation in the average size of the classical orbit of the order of
\begin{gather*}
\biggl( \!\frac{\Delta a}{a} \!\biggr)_{\!\!\text{c}} \approx \frac{3GM}{c^2 a} ( 1+ \tfrac{5}{2}e^2 )
\end{gather*}
e.g. $\sim 6\times 10^{-9}$ for Jupiter and smaller for the outer planets;  it is anyway smaller than the present accuracy in $a$ and therefore can be neglected.''
\end{quote}
(Therein, $a$ refers to the Keplerian semimajor axis.)  It is possible that future experiments will be sensitive to this first-order relativistic correction to the size of orbits.

\begin{figure}
\centering
\includegraphics[width=\columnwidth]{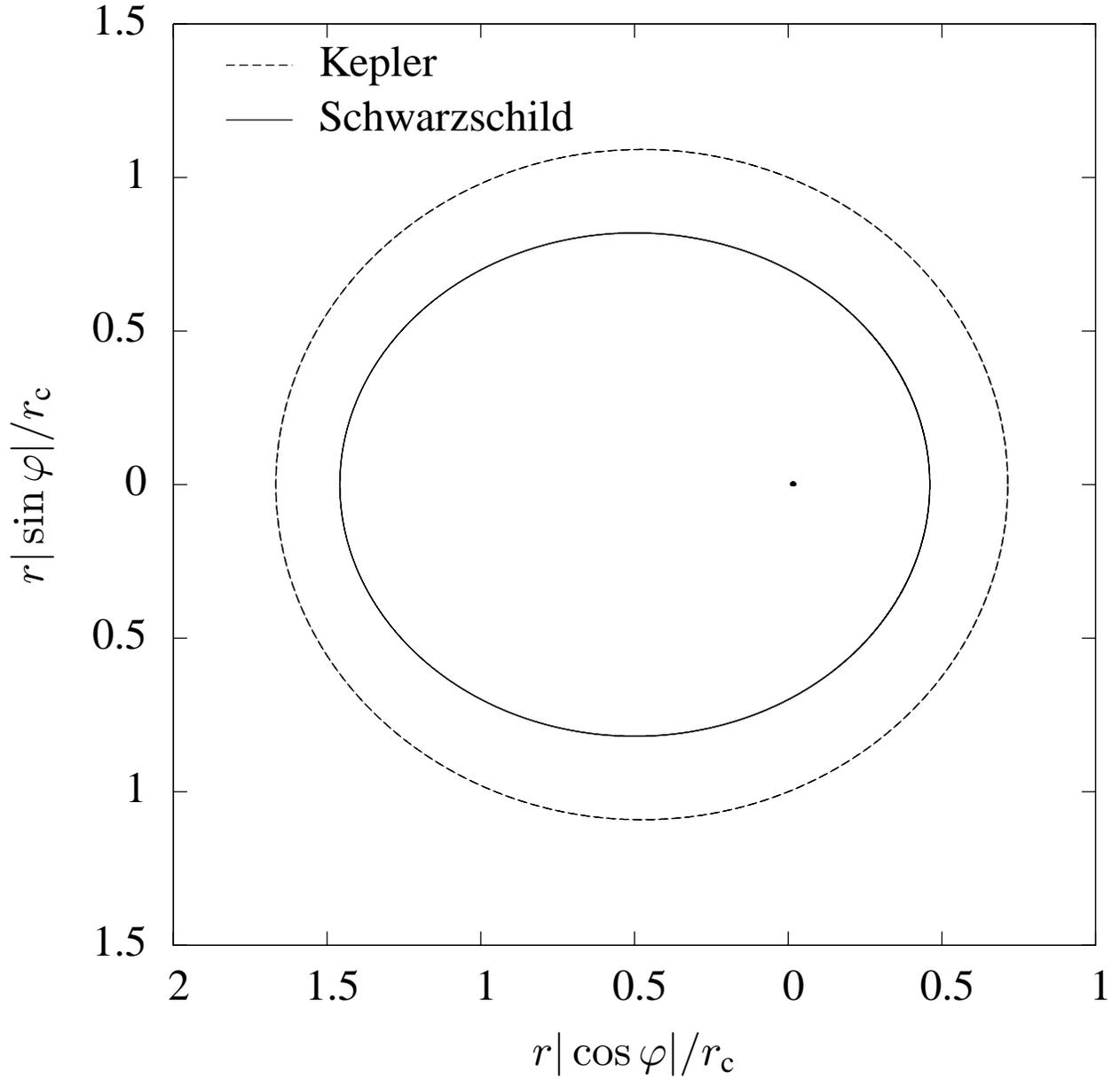}
\caption{\label{fig_reduced} Comparison of Schwarzschild orbit (\ref{eq_S_eoo_concise}) to a corresponding Keplerian orbit (\ref{eq_N_eoo}).  Reduced radial coordinate is one orbital characteristic due to relativity.  The semi-minor axis is reduced more than the semi-major axis, so that the Schwarzschild orbit is also more eccentric than the corresponding Keplerian orbit.  These effects are exaggerated by the choice of parameters $(e=0.4,\;\epm=0.1)$ for purposes of illustration, and precession has been removed $(\tilde{\kappa} \rightarrow 1)$ in order to emphasize the size and shape of the Schwarzschild orbit.  However, these effects are present for smaller (non-zero), reasonably chosen values of $e$ and $\epm$ as well.}
\end{figure}

These characteristics are illustrated in Fig.~\ref{fig_reduced}, in which orbits derived from the Schwarzschild geometry (\ref{eq_S_eoo_concise}) and Newtonian mechanics (\ref{eq_N_eoo}) are compared.  These characteristics are exaggerated by the choice of parameters $(e=0.4,\;\epm=0.1)$ for purposes of illustration, and precession has been removed $(\tilde{\kappa} \rightarrow 1)$ in order to emphasize the size and shape of the Schwarzschild orbit.  However, Schwarzschild orbits have reduced radial coordinates for smaller (non-zero), reasonably chosen values of $e$ and $\epm$ as well;  Figure~\ref{fig_reduced} correctly illustrates that Schwarzschild orbits are smaller than corresponding Keplerian orbits.

When compared to Keplerian orbits, the Schwarzschild geometry predicts orbits that are more eccentric,
\begin{gather}
\delta \tilde{e} \equiv \tilde{e} - e \approx 3\epm e. \label{eq_S_de}
\end{gather}
This characteristic is also displayed in Fig.~\ref{fig_reduced}, in which it is noticeable that the semi-minor axis is reduced more than the semi-major axis.  This quantitative prediction may be subjected to observational tests.  It is possible that part of the discrepency between observed and calculated eccentricities in many astrophysical systems may be accounted for by this contribution.  Bosch {et~al.} \cite{bosch} find galaxy and cluster substructure eccentricity distributions to be strongly skewed toward high eccentricity.  Also, the discrepancy in the eccentricity of the massive binary blackhole system in OJ~287 is found \cite{valtonen} to be of the order expressed by Eq.~(\ref{eq_S_de}).  Champion {et~al.} \cite{champion} have identified an eccentric binary millisecond pulsar (PSR~J1903+0327) ``...that requires a different formation mechanism...'' in order to reconcile the large eccentricity and short spin period.

%%%%%%%%%%%%%%%%%%%%%%%%%
\subsection{Schwarzschild Energy Parameters}\label{SubSec_S_Energy}
%%%%%%%%%%%%%%%%%%%%%%%%%  

For a particular Keplerian orbit, identified by total energy, it is necessary to identify the corresponding Schwarzschild energy for an orbit described by the same angular momentum.  This provides a relativistic correction to Newtonian energies for Keplerian orbits.  This is also useful when comparing orbital properties using energy diagrams.  The generalized virial theorem \cite{ChaCon,bonazzola,vilain,gourgoulhon} provides this relation for circular Schwarzschild orbits, and is easily extended to more general metrics.  Referring to (\ref{eq_S_eom}) and (\ref{eq_S_eff_pot}), a Schwarzschild potential energy parameter for a circular orbit is defined as
\begin{gather}
\tilde{V}_\text{c} \equiv -\frac{GM}{\tilde{r}_\text{c}}-\epm \frac{GMr^2_\text{c}}{\tilde{r}^3_\text{c}},
\end{gather}
where $\tilde{r}_\text{c}$ is the radius of a relativistic circular orbit (\ref{eq_S_1st_coeff_r}), and $\epm$ is defined in Eq.~(\ref{eq_matter}).  According to the virial theorem,
\begin{gather}
\tilde{T}_\text{c} = \frac{GM}{2\tilde{r}_\text{c}} + 3\epm \frac{GMr^2_\text{c}}{2\tilde{r}^3_\text{c}},
\end{gather}
where $\tilde{T}_\text{c}$ is a Schwarzschild kinetic energy parameter.  Therefore, using $\tilde{E}_\text{c} \equiv \tilde{T}_\text{c} + \tilde{V}_\text{c}$,
\begin{align}
\tilde{E}_\text{c} &= - \frac{GM}{2\tilde{r}_\text{c}} + \epm \frac{GMr^2_\text{c}}{2\tilde{r}^3_\text{c}}\\
&= - \frac{GM}{2r_\text{c}} \frac{r_\text{c}}{\tilde{r}_\text{c}} \biggl( 1 - \epm \frac{r^2_\text{c}}{\tilde{r}^2_\text{c}} \biggr) \label{eq_S_Vir_Energy}\\
&\approx E_\text{c}(1+2\epm), \label{eq_S_1st_Vir_Energy}
\end{align}
where Eq.~(\ref{eq_S_1st_coeff_r}) is used in the last step, and $E_\text{c} = -GM/2r_\text{c}$ is the total energy per unit mass for a circular Keplerian orbit of radius $r_\text{c}$.  This may instead be expressed as
\begin{gather}
\tilde{E}_\text{c} \approx E_\text{c} - 2\epm|E_\text{c}|. \label{eq_S_1st_Energy_circ}
\end{gather}
Therefore, $\delta \tilde{E}_\text{c} \approx -2\epm|E_\text{c}| < 0$;  The energy of a circular Schwarzschild orbit $(\tilde{e} = e = 0)$ is less than that of a corresponding Keplerian orbit.  This relation is displayed in Fig.~\ref{fig_dE_circular}.

More generally, the quantitative prediction of increased eccentricity provides an avenue by which Keplerian and Schwarzschild energies may be compared.  The total energy per unit mass for a Keplerian orbit may be expressed in terms of the eccentricity as \cite{goldstein,MT}
\begin{gather}
E/E_\text{c} = 1 - e^2. \label{eq_Kep_Energy}
\end{gather}
Analogously, an approximate Schwarzschild energy parameter is defined by ansatz to be
\begin{gather}
\tilde{E}/\tilde{E}_\text{c} \approx 1 - \tilde{e}^2. \label{eq_ansatz}
\end{gather}
The Schwarzschild energy parameter (\ref{eq_ansatz}) is then expressed in terms of the Keplerian energy using Eqs.~(\ref{eq_S_1st_coeff_e}), (\ref{eq_S_1st_Vir_Energy}) and (\ref{eq_Kep_Energy}),
\begin{gather}
\tilde{E} \approx E - 2\epm(1 - 4e^2)|E_\text{c}|. \label{eq_S_1st_Energy}
\end{gather}
(This ansatz is investigated in Sec.~\ref{SubSec_S_SC_energy}, wherein (\ref{eq_S_1st_Energy}) is shown to approximate well a more detailed parameterization and is a good approximation for near-circular orbits.)  Therefore, $\delta \tilde{E} \approx - 2\epm|E_\text{c}|(1-4e^2) < 0$;  The Schwarzschild energy is less than the corresponding Keplerian energy until the energies become equal for $e_0 \approx \tfrac{1}{2}$.  The Schwarzschild energy parameter becomes greater than the corresponding Keplerian energy for $e>e_0$.  For near-circular orbits the Schwarzschild energy parameter always lies below the Keplerian energy.  This relation is displayed in Fig.~\ref{fig_dE}.  The value of eccentricity for which the Schwarzschild energy parameter is equal to the corresponding Keplerian energy is approximately the same as that for which the apocenter distances are equal, $e_0 \approx e_{+} \approx 1/2$. (See Sec.~\ref{SubSec_S_Char}.)  However, this approximate energy parametrization is not expected to be accurate for values of $e$ approaching $1/2$.  A more accurate treatment in Secs.~\ref{SubSec_S_SC_char}~and~\ref{SubSec_S_SC_energy} results in $e_{+} < e_0$, as expected.

%%%%%%%%%%%%%%%%%%%%%%%%%%
\section{Self-Consistent Keplerian Limit}\label{Sec_S_self_cons}
%%%%%%%%%%%%%%%%%%%%%%%%%%

The relativistic equation of orbit (\ref{eq_S_eoo_concise}) is useful for describing characteristics of relativistic orbits in a Keplerian limit.  However, the solution is inaccurate in describing long-term orbital behavior.  It is easy to verify that (\ref{eq_S_eoo}) solves (\ref{eq_S_eoo_DE}) only to first-order in $\epm$.  The Schwarzschild equation of orbit (\ref{eq_S_eoo}) may instead be expressed as
\begin{gather}
\frac{r_\text{c}}{r} = \frac{r_\text{c}}{\tilde{r}_\text{c}} ( 1 + \tilde{e} \cos{\tilde{\kappa}\varphi}), \label{eq_S_eoo_alt}
\end{gather}
where $\tilde{r}_\text{c}$, $\tilde{e}$, and $\tilde{\kappa}$ are defined in Eqs.~(\ref{eq_S_coeff_r})--(\ref{eq_corr_princ}).  Substituting (\ref{eq_S_eoo_alt}) into the equation of motion (\ref{eq_S_eoo_DE}) and discarding the term nonlinear in eccentricity results in the two conditions:
\begin{align}
1 - \tilde{\kappa}^2 &\approx 6\epm ( \tilde{r}_\text{c}/r_\text{c} )^{-1}; \label{eq_cond1}\\
( \tilde{r}_\text{c}/r_\text{c} )^{-1} &\approx 1+3\epm ( \tilde{r}_\text{c}/r_\text{c} )^{-2}. \label{eq_cond2}
\end{align}
The first condition (\ref{eq_cond1}) is satisfied only to first order in $\epm$, and the second condition (\ref{eq_cond2}) is satisfied only to second order in $\epm$.  Fortunately, consistency is achieved by taking $\tilde{r}_\text{c}$ and $\tilde{\kappa}$ to be independent functions of $\epm$ and making the following replacements in Eq.~(\ref{eq_S_eoo_alt}):
\begin{align}
\tilde{\kappa} &\rightarrow (\tilde{\kappa}^2 - \kappa_0^2)^\frac{1}{2}; \label{eq_replace1}\\
\tilde{r}_\text{c}/r_\text{c} &\rightarrow \tilde{r}_\text{c}/r_\text{c} - \tfrac{1}{2} \lambda_0^3. \label{eq_replace2}
\end{align}
The constants $\kappa_0^2$ and $\lambda_0^3$ are then found by substituting Eq.~(\ref{eq_S_eoo_alt}) into the equation of motion (\ref{eq_S_eoo_DE}), resulting in
\begin{align}
\kappa_0^2 &= 1 - 6 \epm - (1 - 12 \epm)^\frac{1}{2} = 18 \epmsquared + \mathcal{O} (\epmcubed),\\
\lambda_0^3 &= \frac{1-9\epm}{1-3\epm} - (1 - 12 \epm)^\frac{1}{2} = 54 \epmcubed + \mathcal{O} (\epmfourth). \label{eq_constant2}
\end{align}
A self-consistent equation of orbit may now be expressed as
\begin{gather}
\frac{\tilde{r}_\text{c}}{r} = 1 + \tilde{e}\cos{\tilde{\kappa}\varphi}, \label{eq_S_SC_eoo}
\end{gather}
with the following definitions:
\begin{align}
\tilde{r}_\text{c}/r_\text{c} &\equiv \tfrac{1}{2} + \tfrac{1}{2} \sqrt{1-12\epm}; \label{eq_S_SC_coeff_r}\\
\tilde{e}/e &\equiv \frac{1}{1-3\epm}; \label{eq_S_SC_coeff_e}\\
\tilde{\kappa} &\equiv (1-12\epm)^\frac{1}{4}. \label{eq_S_SC_coeff_phi}
\end{align}
This solution is consistent to all orders in $\epm$, and the more approximate orbital equation (\ref{eq_S_eoo_concise}) of Sec.~\ref{Sec_S_KepLim} is recovered by expanding (\ref{eq_S_SC_coeff_r})--(\ref{eq_S_SC_coeff_phi}) and keeping terms first order in $\epm$:
\begin{align}
\tilde{r}_\text{c} &\approx r_\text{c}(1 - 3\epm - 9\epmsquared - 54\epmcubed);\\
\tilde{e} &\approx e(1 + 3\epm + 9\epmsquared + 27\epmcubed); \label{eq_S_SC_eccen}\\
\tilde{\kappa} &\approx 1 - 3\epm - \tfrac{27}{2}\epmsquared - \tfrac{189}{2}\epmcubed.
\end{align}
Long-term behavior is predicted very accurately using the exact expressions (\ref{eq_S_SC_coeff_r})--(\ref{eq_S_SC_coeff_phi}).  The radius of relativistic circular orbit (\ref{eq_S_SC_coeff_r}) is identical to the result obtained by minimizing the effective potential (\ref{eq_S_rc_eff_pot}).

A systematic verification may be carried out by direct substitution of Eq.~(\ref{eq_S_SC_eoo}) into Eq.~(\ref{eq_S_eoo_DE}) using the exact definitions (\ref{eq_S_SC_coeff_r})--(\ref{eq_S_SC_coeff_phi}) for $\tilde{r}_\text{c}$, $\tilde{e}$, and $\tilde{\kappa}$, and discarding the term nonlinear in eccentricity.  However, the justification for discarding the term nonlinear in eccentricity is the correspondence principle.  Arguments concerning which terms to keep based only on a direct comparison of the relative magnitudes of higher-order and lower-order terms lead to contradictions.  Rather, the domain of validity is expressed by subjecting the solution (\ref{eq_S_SC_eoo}) to condition (\ref{eq_linearize}).  Evaluating the equation of orbit (\ref{eq_S_SC_eoo}) at pericenter results in
\begin{gather}
\frac{r_\text{c}}{\tilde{r}_{-}} = \frac{1+\tilde{e}}{\tilde{r}_\text{c}/r_\text{c}}.
\end{gather}
Therefore, according to (\ref{eq_linearize}), the domain of validity is given by
\begin{gather}
\tilde{e}/e + 2 \bigl(1 - \tilde{r}_\text{c}/r_\text{c} \bigr) \ll 1, \label{eq_S_SC_valid}
\end{gather}
Where $\tilde{r}_\text{c}/r_\text{c}$ and $\tilde{e}$ are given by (\ref{eq_S_SC_coeff_r}) and (\ref{eq_S_SC_coeff_e}), respectively.  This condition is consistent with Eq.~(\ref{eq_S_1st_valid}) to first order in $\epm$. Of course, the further condition $\epm \le 1/12$ is also necessary.

It is worth mention that this self-consistency program is independent of the eccentricity.  A replacement of the form $\tilde{e} \rightarrow \tilde{e} + \eta_0$, such as in (\ref{eq_replace1}) and (\ref{eq_replace2}), may be imposed, but $\eta_0$ cannot be determined by insisting that the solution (\ref{eq_S_eoo_alt}) satisfy the equation of motion (\ref{eq_S_eoo_DE}).  No adjustments are made to further increase the accuracy of the relativistic eccentricity.  The accuracy of the relativistic eccentricity is verified only by agreements of (\ref{eq_S_SC_eoo}) with exact numerical solutions of the equation of orbit (\ref{eq_S_eoo_DE}).

The self-consistent equation of orbit (\ref{eq_S_SC_eoo}) is in agreement with exact numerical solutions of (\ref{eq_S_eoo_DE}) for a large parameter space, including high-eccentricity $(1 > e \gg 0)$ and very relativistic environments $(1/12 > \epm \gg 0)$.   These periodic solutions are compared using relative errors in radial coordinate and angular frequency separately.  The relative error in radial coordinate is defined as
\begin{gather}
\delta r/r \equiv \frac{r/r_\text{c} - \tilde{r}/r_\text{c}}{r/r_\text{c}}, \label{eq_rel_err_r}
\end{gather}
where $r/r_\text{c}$ is the exact numerical solution of (\ref{eq_S_eoo_DE}), and $\tilde{r}/r_\text{c}$ is the self-consistent solution (\ref{eq_S_SC_eoo}).  For this purpose solutions are compared over a small number or periods.  The relative error in angular frequency is defined as
\begin{gather}
\delta\kappa/\kappa \equiv 1- \varphi_\text{n}/\tilde{\varphi}_\text{n}, \label{eq_rel_err_phi}
\end{gather}
where $\kappa\varphi_\text{n} \equiv 2\pi\text{n}$ is the phase of the numerical solution over $\text{n}$ cycles, and $\tilde{\kappa}\tilde{\varphi}_\text{n} = 2\pi\text{n}$ is the phase of the self-consistent solution over the same number of cycles.  This error is determined using $\text{n} = 1600$ cycles, and in each case $\varphi_\text{n} - \tilde{\varphi}_\text{n}$ represents the total phase difference.

\begin{figure}
\centering
\includegraphics[width=\columnwidth]{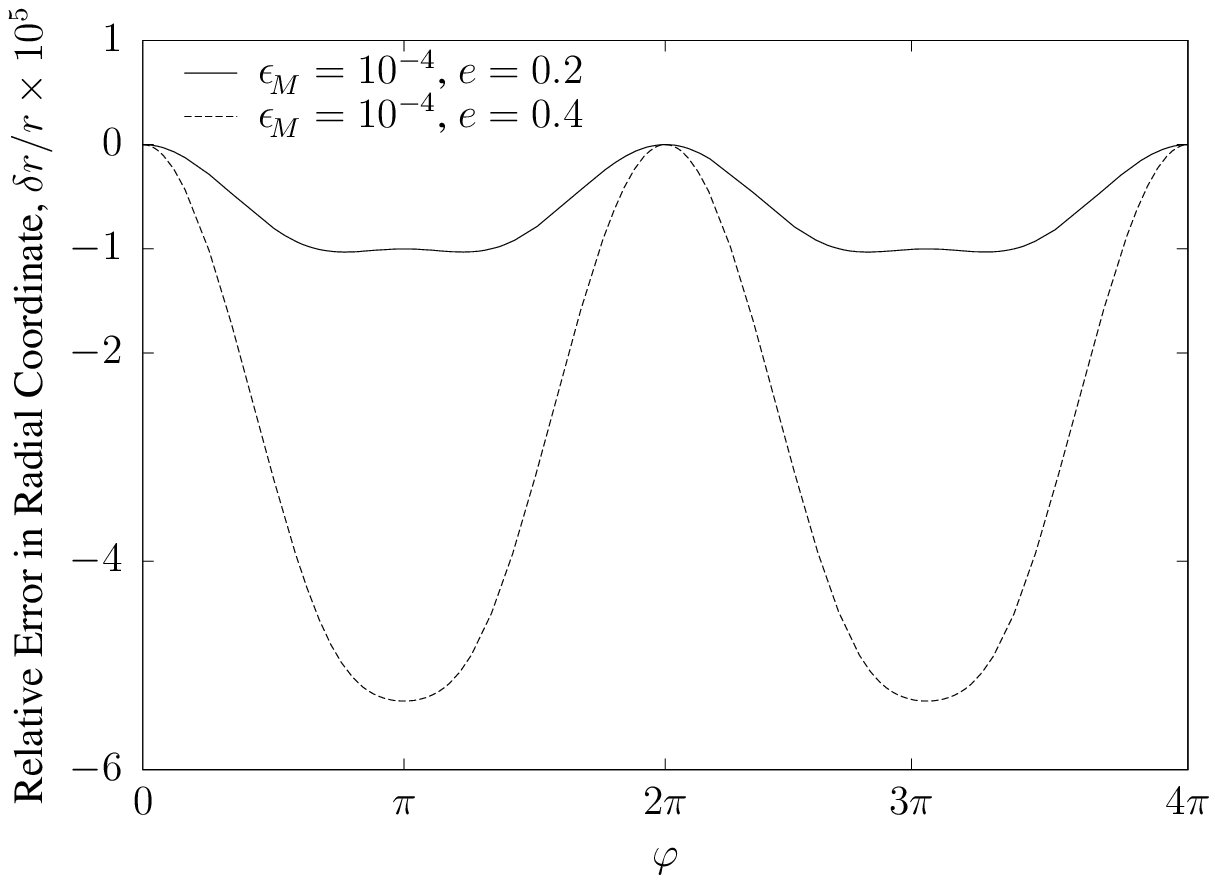}
\caption{\label{fig_intermediate} Comparison of self-consistent Schwarzschild solution (\ref{eq_S_SC_eoo}) to the exact numerical solution of (\ref{eq_S_eoo_DE}) for intermediate values of $\epm$ and $e$.  The relative error, Eq.~(\ref{eq_rel_err_r}), is plotted for the initial interval $0 \le\varphi\le 4\pi$.  The phases are nearly equal over this initial small interval, so that this figure effectively represents the relative error in the radial coordinate:  $\delta r/r \sim 10^{-5}$.  In each of these two cases the relative error in angular frequency, Eq.~(\ref{eq_rel_err_phi}), is found to be: $\delta 
\kappa/\kappa \sim 10^{-8}$.}
\end{figure}

\begin{figure}
\centering
\includegraphics[width=\columnwidth]{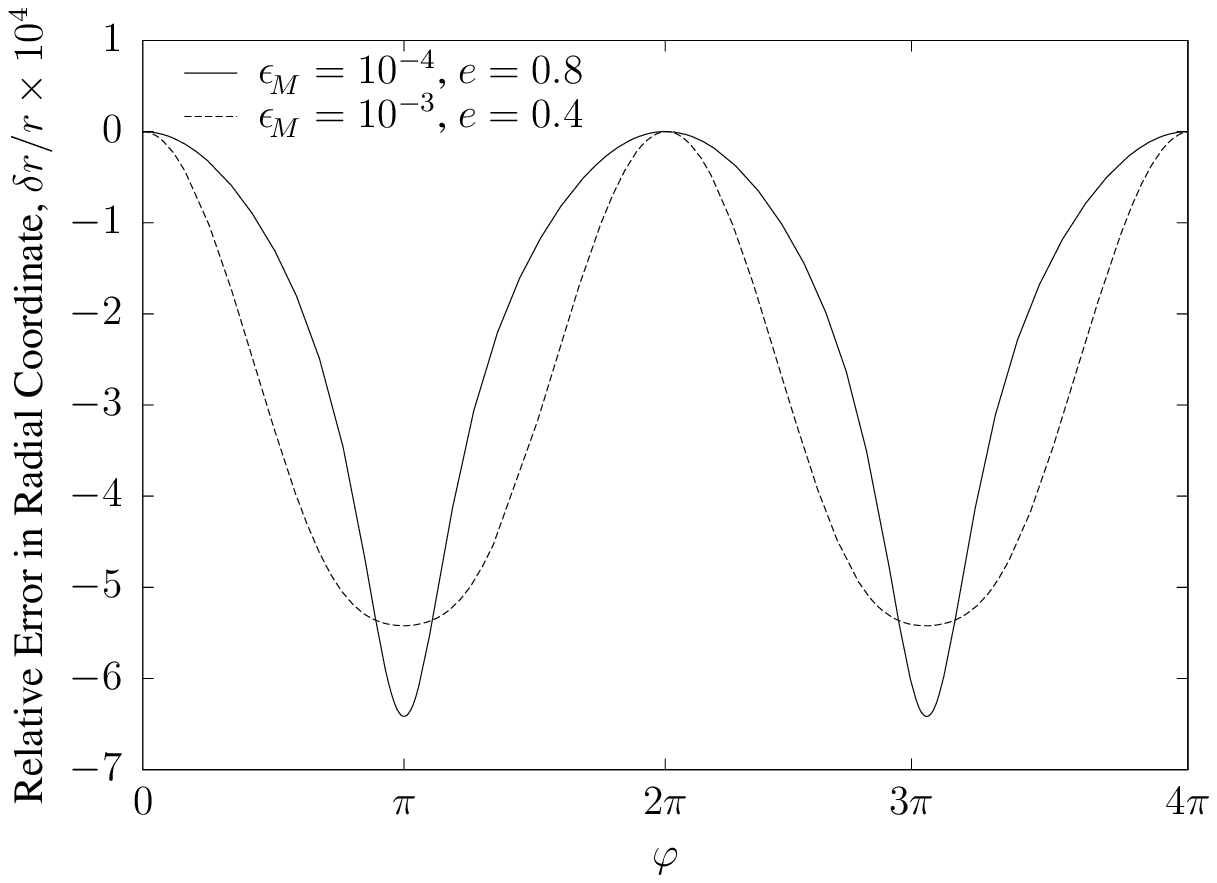}
\caption{\label{fig_extreme} Comparison of self-consistent Schwarzschild solution (\ref{eq_S_SC_eoo}) to the exact numerical solution of (\ref{eq_S_eoo_DE}) for more extreme values of $\epm$ and $e$.  The relative error, Eq.~(\ref{eq_rel_err_r}), is plotted for the initial interval \hbox{$0 \le\varphi\le 4\pi$.}  The phases are nearly equal over this initial small interval, so that this figure effectively represents the relative error in the radial coordinate:  $\delta r/r \sim 10^{-4}$.  The relative error in angular frequency, Eq.~(\ref{eq_rel_err_phi}), is, for each of these two cases: $\delta 
\kappa/\kappa \sim 10^{-8}$ for $\epm=10^{-4}$ and $e=0.8$; and $\delta 
\kappa/\kappa \sim 10^{-6}$ for $\epm=10^{-3}$ and $e=0.4$.}
\end{figure}

Errors for several cases, including relatively large values of $\epm$ and $e$, are plotted in \hbox{Figs.~\ref{fig_moderate}-\ref{fig_extreme}}.  For reference: $\epm\sim 10^{-8}$ for the earth-sun system $(r_\text{c}\sim 10^{8} \rm/2)$; $\epm\sim 10^{-6}$ for an outlying $(r_\text{c}\sim 1\,\text{Mpc})$ giant spiral galaxy orbiting a cluster containing one thousand giant spiral galaxies $(r_\text{c}\sim 10^{6} \rm/2)$;  $\epm = 10^{-3}$ for a star orbiting a super-massive blackhole at a distance corresponding to $r_\text{c}=500\rm$.  It is worth noting that in every case the largest error is near apocenter $(\varphi = \pi)$, and that this error is in favor of the stated characteristic of reduced apocenter.  The exact numerical value for the radial coordinate is smaller than that determined by the equation of orbit derived in this Keplerian limit. 

This self-consistent solution (\ref{eq_S_SC_eoo}) is identical in form to the more approximate solution (\ref{eq_S_eoo_concise}) of Sec.~\ref{Sec_S_KepLim}.  However, the coefficients (\ref{eq_S_SC_coeff_r})--(\ref{eq_S_SC_coeff_phi}) result in an equation of orbit that accurately predicts long-term behavior.  The more approximate solution of Sec.~\ref{Sec_S_KepLim} is a limiting case of this self-consistent solution, lending value to the simpler approach and first-order predictions.

%%%%%%%%%%%%%%%%%%%%%%%%%
\subsection{Characteristics of Schwarzschild Orbits}\label{SubSec_S_SC_char}
%%%%%%%%%%%%%%%%%%%%%%%%%

The characteristics of Schwarzschild orbits are expressed more accurately using the self-consistent equation of orbit (\ref{eq_S_SC_eoo})--(\ref{eq_S_SC_coeff_phi}).  The self-consistent equation of orbit predicts the rate of precession to be
\begin{align}
(2\pi)^{-1}\Delta\varphi &= \tilde{\kappa}^{-1} - 1,\\
&=3\epm + \tfrac{45}{2}\epmsquared + \mathcal{O}(\epmcubed),
\end{align}
The second-order term $\Delta\varphi^{(2)} \sim \tfrac{45}{2}\epmsquared$ is in close agreement with that calculated by Ashby \cite{ashby} $\Delta\varphi^{(2)}_\textsc{a} \sim (\tfrac{27}{2} + \tfrac{3}{4} e^2) \epmsquared$, so that $1.58 < \Delta\varphi^{(2)}/\Delta\varphi^{(2)}_\textsc{a} < 1.67$.

The self-consistent equation of orbit predicts the radius of stable circular orbit to be reduced,
\begin{gather}
\delta \tilde{r}_\text{c}/r_\text{c} = -3\epm - 9\epmsquared - \mathcal{O}(\epmcubed).
\end{gather}
This expression is in agreement with the Schwarzschild geometry;  The radius of circular orbit as predicted by this self-consistent solution (\ref{eq_S_SC_coeff_r}) is identical to that calculated from the Schwarzschild geometry by minimizing the effective potential (\ref{eq_S_rc_eff_pot}).  The relativistic apsides are
\begin{gather}
\tilde{r}_\pm/r_\pm = \frac{\tilde{r}_\text{c}}{r_\text{c}} \frac{1 \mp e}{1 \mp e/(1 - 3\epm)},
\end{gather}
The relativistic apocenter is smaller than the corresponding Keplerian apocenter for values of eccentricity less than
\begin{align}
e_{+} &= \biggl( \! 1 + \frac{\tilde{r}_\text{c}}{r_\text{c}} \frac{\tilde{e}}{e} \biggr)^{\!\!-1}\\
&\approx \tfrac{1}{2}(1 + \tfrac{9}{2}\epmsquared). \label{eq_S_SC_e_apo}
\end{align}
(There is no correction first order in $\epm$ here.)

The self-consistent equation of orbit predicts the eccentricity of a noncircular orbit to be increased,
\begin{align}
\delta \tilde{e}/e &= 3\epm \tilde{e}/e \label{eq_S_SC_de}\\
&= 3\epm + 9\epmsquared + \mathcal{O}(\epmcubed). \label{eq_S_SC_de_Taylor}
\end{align}

%%%%%%%%%%%%%%%%%%%%%%%%%
\subsection{Schwarzschild Energy Parameters}\label{SubSec_S_SC_energy}
%%%%%%%%%%%%%%%%%%%%%%%%%

A more accurate relation between the Schwarzschild energy parameter for a circular orbit and that derived from Newtonian mechanics is constructed using the virial theorem together with the radius of circular orbit as derived from the self-consistent equation of orbit.  (See Sec.~\ref{SubSec_S_Energy}.)  The Schwarzschild energy parameter is given in terms of the Keplerian energy by Eqs.~(\ref{eq_S_Vir_Energy})~and~(\ref{eq_S_SC_coeff_r}),
\begin{align}
\tilde{E}_\text{c}/E_\text{c} &= \frac{\tilde{r}_\text{c}/r_\text{c} - 4\epm}{(\tilde{r}_\text{c}/r_\text{c})^3} \label{eq_S_Vir_Energy_exact}\\
&= 1 + 2\epm + 9\epmsquared + 54\epmcubed + \mathcal{O}(\epmfourth).
\end{align}
This result is also obtained by substituting the radius of circular orbit (\ref{eq_S_rc_eff_pot}) into the effective potential (\ref{eq_S_eff_pot}), and is in agreement with the approximate relation (\ref{eq_S_1st_Vir_Energy}) to first order in $\epm$.

More generally, a Schwarzschild energy parameter for noncircular, approximately Keplerian orbits is defined by
\begin{gather}
\tilde{e} \equiv \frac{R_+ - R_-}{R_+ + R_-} \approx \frac{\tilde{r}_+ - \tilde{r}_-}{\tilde{r}_+ + \tilde{r}_-},
\end{gather}
where $R_\pm=R_\pm (\tilde{E})$ are the relativistic apsides as determined from the intersection of the effective potential (\ref{eq_S_eff_pot}) with a line of constant energy $\tilde{E}$.  That is,
\begin{gather}
r^2_\text{c}\ell^{-2}\tilde{E} = - \frac{r_\text{c}}{R_\pm} + \frac{1}{2}\frac{r^2_\text{c}}{R_\pm^2} - \epm \frac{r^3_\text{c}}{R_\pm^3}.  \label{eq_S_cubic}
\end{gather}
The solution is then inverted, yielding $\tilde{E}(\tilde{e})$.  The cubic (\ref{eq_S_cubic}) is solved using Vi\`{e}te's (1540--1603) trick \cite{fulling},
\begin{gather}
\frac{r_\text{c}}{R_\pm} = \frac{1}{6\epm} \bigl[ 1 + 2 \beta \cos{( \phi_\pm + \tfrac{1}{3} \arccos{\gamma})} \bigr],
\end{gather}
where $\phi_\pm = \pi \mp \pi/3$, and the following functions are defined:
\begin{align}
\beta &\equiv \sqrt{1-12\epm};\\
\gamma &\equiv \beta^{-3} [ 1 - 18\epm ( 1 - 3\epm \tilde{E}/E_\text{c} ) ].
\end{align}
(Notice that $2r_\text{c}^2\ell^{-2} = -1/E_\text{c}$.)  Therefore, the relativistic eccentricity is given by
\begin{gather}
\tilde{e} = \frac{\sqrt{3} \beta \sin{ (\tfrac{1}{3} \arccos{ \gamma }) }}{ 1 - \beta \cos{(\tfrac{1}{3} \arccos{\gamma})} }. \label{eq_S_e_exact}
\end{gather}
This is verified by setting $\tilde{e}=0$ and calculating the energy of a circular Schwarzschild orbit, resulting in Eq.~(\ref{eq_S_Vir_Energy_exact}).  Inverting (\ref{eq_S_e_exact}) results in an expression for the Schwarzschild energy parameter,
\begin{gather}
\tilde{E}/E_\text{c} = \frac{1}{3\epm} \biggl\{ 1 - \frac{1}{18\epm} [ 1 - \beta^3\cos{(3\arccos{\lambda})} ] \biggr\}, \label{eq_S_Energy_exact}
\end{gather}
where
\begin{align}
\lambda &= \beta^{-1} \zeta^{-1} ( 1 + \sqrt{1 + \xi\zeta} ),  \\
\zeta   &= \frac{3 + \tilde{e}^2}{\tilde{e}^2}, \\
\xi	&= \frac{3\beta^2 - \tilde{e}^2}{\tilde{e}^2}.
\end{align}
Dividing (\ref{eq_S_Energy_exact}) by (\ref{eq_S_Vir_Energy_exact}) and expanding in powers of $\epm$ results in an expression comparable to the earlier energy parameterization ansatz (\ref{eq_ansatz}),
\begin{gather}\begin{split}
\tilde{E}/\tilde{E}_\text{c} ={} & ( 1 - \tilde{e}^2 ) [ 1 + 2\tilde{e}^2 \epm + \tilde{e}^2 ( 14 + 5 \tilde{e}^2 ) \epmsquared \\
& + \mathcal{O}(\tilde{e}^2 \epmcubed) ].\end{split} \label{eq_S_compare_to_ansatz}
\end{gather}
In the limit $\epm \rightarrow 0$ the Keplerian result (\ref{eq_Kep_Energy}) is recovered.  The Schwarzschild energy parameter may instead be expressed in terms of the energy and eccentricity of the  corresponding Keplerian orbit using (\ref{eq_S_Energy_exact}) together with (\ref{eq_S_SC_coeff_e}) and (\ref{eq_Kep_Energy}),
\begin{gather}\begin{split}
\tilde{E}/E_\text{c} ={} & E/E_\text{c} + 2\epm ( 1 - 3e^2 - e^4 ) \\
& + \epmsquared ( 9 - 18e^2 - 37e^4 - 5e^6 ) + \mathcal{O}(\epmcubed). \end{split} \label{eq_S_Energy_exact_Taylor}
\end{gather}
Neglecting terms of orders $\epm e^4$\!, $\epmsquared$, and smaller in Eq.~(\ref{eq_S_Energy_exact_Taylor}) results in
\begin{gather}
\tilde{E} \approx E - 2\epm(1-3e^2)|E_\text{c}|.
\end{gather}
This result is similar to the energy relation (\ref{eq_S_1st_Energy}) derived from the simple ansatz (\ref{eq_ansatz}), lending value to the much simpler approach of Sec.~\ref{SubSec_S_Energy}.  Corrections to the orginal ansatz (\ref{eq_ansatz}) are given by Eq.~(\ref{eq_S_compare_to_ansatz}).  The orginal ansatz (\ref{eq_ansatz}) and the resulting relation between the Schwarzschild energy parameter and corresponding Keplerian energy (\ref{eq_S_1st_Energy}) approximate well this more detailed parameterization, (\ref{eq_S_compare_to_ansatz}) and (\ref{eq_S_Energy_exact_Taylor}), for near-circular orbits requiring small relativistic corrections.

Referring to (\ref{eq_S_Energy_exact_Taylor}), the energy of a Schwarzschild orbit is smaller than that for the corresponding Keplerian orbit, until the energies become equal for
\begin{align}
1 - 3e_0^2 - e_0^4 \approx 0,
\end{align}
so that $e_0 \approx 0.55$.  The relativistic energy parameter becomes greater than the corresponding Keplerian energy for $e>e_0$.  This value of eccentricity is larger than that for which the corresponding apocenter distances are equal, as determined from Eq.~(\ref{eq_S_SC_e_apo}); for $\epm = 1/12$, $e_+ \approx 0.52$, so that $e_{+}<e_0$ for all allowed values of $\epm$.  (Including both the $\epm$- and $\epmsquared$-terms in Eq.~(\ref{eq_S_Energy_exact_Taylor}) results in $e_0 \approx 0.550 + 1.99\times10^{-2}\epm$, and the relation $e_{+}<e_0$ holds.)  This is expected since it is required that $\tilde{r}_{+}>r_{+}$ for any energy $\tilde{E}=E>E_\text{c}$;  Referring to an energy diagram, $\tilde{E}$ must lie below $E$ in order for $\tilde{r}_{+}=r_{+}$.

%%%%%%%%%%%%%%%%%%%%%%%%%
\section{Reissner-Nordstr\"{o}m Orbits in Keplerian Limit}\label{Sec_RN_KepLim}
%%%%%%%%%%%%%%%%%%%%%%%%%

The path of a small, electrically neutral test mass near a spherically-symmetric mass $M$ with charge $Q$ is described by the Reissner-Nordstr\"{o}m geometry \cite{MTW,PleKra,OR,dinverno,carroll2,HobEfsLas,chaliasos,TP,weyl0,reissner,nordstrom}.  The metric is given by Eq.~(\ref{eq_metric_general}) with
\begin{gather}
\mathrm{e}^{2\nu(r)} = \mathrm{e}^{2\mu(r)} = 1 - \rm/r + \rqsquared/r^2,
\end{gather}
where $\rqsquared \equiv GQ^2/(4\pi\varepsilon_0 c^4)$, and $\varepsilon_0$ is the electrical permittivity of the vacuum.  A far exterior solution is described in the Schwarzschild limit, so that $r\gg \rm>\rq$.  Therefore, the singularities $\mathrm{e}^{2\mu(r^{\vphantom{0}}_{\!0})} = 0$ are irrelevant in the present context.  The equations of motion may be expressed as
\begin{align}
\ell &= r^2\dot\varphi,\\
k &= (1 - \rm/r + \rqsquared/r^2)\,\dot{t},\\
\tfrac{1}{2}(k^2-1)c^2 &=\tfrac{1}{2}\dot{r}^2 + \tilde{V}_\text{eff}, \label{eq_RN_eom}
\end{align}
where an effective potential is defined as
\begin{gather}
r^2_\text{c}\ell^{-2}\tilde{V}_\text{eff} \!\equiv\! - \frac{r_\text{c}}{r} + \frac{1}{2}\frac{r^2_\text{c}}{r^2} - \epm \frac{r^3_\text{c}}{r^3} + \epq\frac{r^2_\text{c}}{r^2} +  \epm\epq\frac{r^4_\text{c}}{r^4}.  \label{eq_RN_eff_pot}
\end{gather}
The parameter $r_\text{c}$ is defined in Sec.~\ref{Sec_S_KepLim} after Eq.~(\ref{eq_S_eff_pot}), and the charge-related relativistic correction parameter is defined as
\begin{gather}
\epq\equiv \frac{\rqsquared}{\rm r_\text{c}} = 2\epm \frac{\rqsquared}{\rmsquared}. \label{eq_epsilonQ}
\end{gather}
The condition $\rq < \rm$ is equivalent to $\epq < 2\epm$, so that for the extreme case in which $\rq=\rm$, $\epq=2\epm$.  However, as discussed below, a consistent treatment of the Reissner-Nordstr\"{o}m (RN) geometry in this Keplerian limit requires the condition $\epq/\epm < \tfrac{3}{2}$.  An equation for the trajectory of a test particle is obtained by eliminating time from Eq.~(\ref{eq_RN_eom}) and differentiating once more with respect to $\varphi$,
\begin{gather}
\frac{\mathrm{d}^2}{\mathrm{d} \varphi^2} \frac{r_\text{c}}{r} + \frac{r_\text{c}}{r} = 1 + 3\epm \Big( \frac{r_\text{c}}{r} \Big)^{\!2} - 2\epq\frac{r_\text{c}}{r} - 4\epm\epq \Big( \frac{r_\text{c}}{r} \Big)^{\!3}\!. \label{eq_RN_eoo_DE}
\end{gather}
Schwarzschild orbits (\ref{eq_S_eoo_DE}) are recovered by setting $\epq=0$, and the conic-sections of Newtonian mechanics (\ref{eq_N_eoo}) are recovered by setting $\epm = \epq=0$.

Reissner-Nordstr\"{o}m orbits in this Keplerian limit are described by neglecting the last terms on the right-hand-sides of Eqs.~(\ref{eq_RN_eff_pot})~and~(\ref{eq_RN_eoo_DE}), resulting in
\begin{gather}
r^2_\text{c}\ell^{-2}\tilde{V}_\text{eff} \!\approx\! - \frac{r_\text{c}}{r} + \frac{1}{2}\frac{r^2_\text{c}}{r^2} - \epm \frac{r^3_\text{c}}{r^3} + \epq\frac{r^2_\text{c}}{r^2}, \label{eq_RNEP2}\\
\frac{\mathrm{d}^2}{\mathrm{d} \varphi^2} \frac{r_\text{c}}{r} + \frac{r_\text{c}}{r} \approx 1 + 3\epm \Big( \frac{r_\text{c}}{r} \Big)^{\!2} - 2\epq\frac{r_\text{c}}{r}. \label{eq_RNeom_phi2}
\end{gather}
The neglected term is only important when describing very close encounters $(r\ll r_\text{c})$.  Following the procedure used to solve Eq.~(\ref{eq_S_eoo_DE}), described by (\ref{eq_linearize})-(\ref{eq_corr_princ}), a first-order solution to (\ref{eq_RNeom_phi2}) may be expressed as
\begin{gather}
\frac{\tilde{r}_\text{c}}{r} \approx 1 + \tilde{e} \cos{\tilde{\kappa}\varphi}, \label{eq_RN_orbits_approx}
\end{gather}
where
\begin{align}
\tilde{r}_\text{c} &\approx r_\text{c} (1 - 3\epm + 2\epq),\\
\tilde{e} &\approx e (1 + 3\epm),\\
\tilde{\kappa} & \approx 1 - 3\epm + \epq. \label{eq_RN_coeff_phi}
\end{align}
A systematic verification may be carried out by substituting (\ref{eq_RN_orbits_approx}) into (\ref{eq_RNeom_phi2}), keeping terms of orders $e$, $\epm$, $\epq$, $e\epm$, and $e\epq$ only.  However, the justification for discarding the term nonlinear in eccentricity is the correspondence principle.  Arguments concerning which terms to discard based only on direct comparisons of relative magnitudes of higher-order and lower-order terms lead to contradictions. Rather, the domain of validity is expressed by subjecting the solution (\ref{eq_RN_orbits_approx}) to condition (\ref{eq_linearize}) at pericenter, resulting in 
\begin{gather}
e(1 + 3\epm) + 2(3\epm - 2\epq) \ll 1. \label{eq_RN_1st_valid}
\end{gather}

%%%%%%%%%%%%%%%%%%%%%%%%%
\subsection{Characteristics of Reissner-Nordstr\"{o}m Orbits}\label{SubSec_RN_char}
%%%%%%%%%%%%%%%%%%%%%%%%%

\begin{figure}
\centering
\includegraphics[width=\columnwidth]{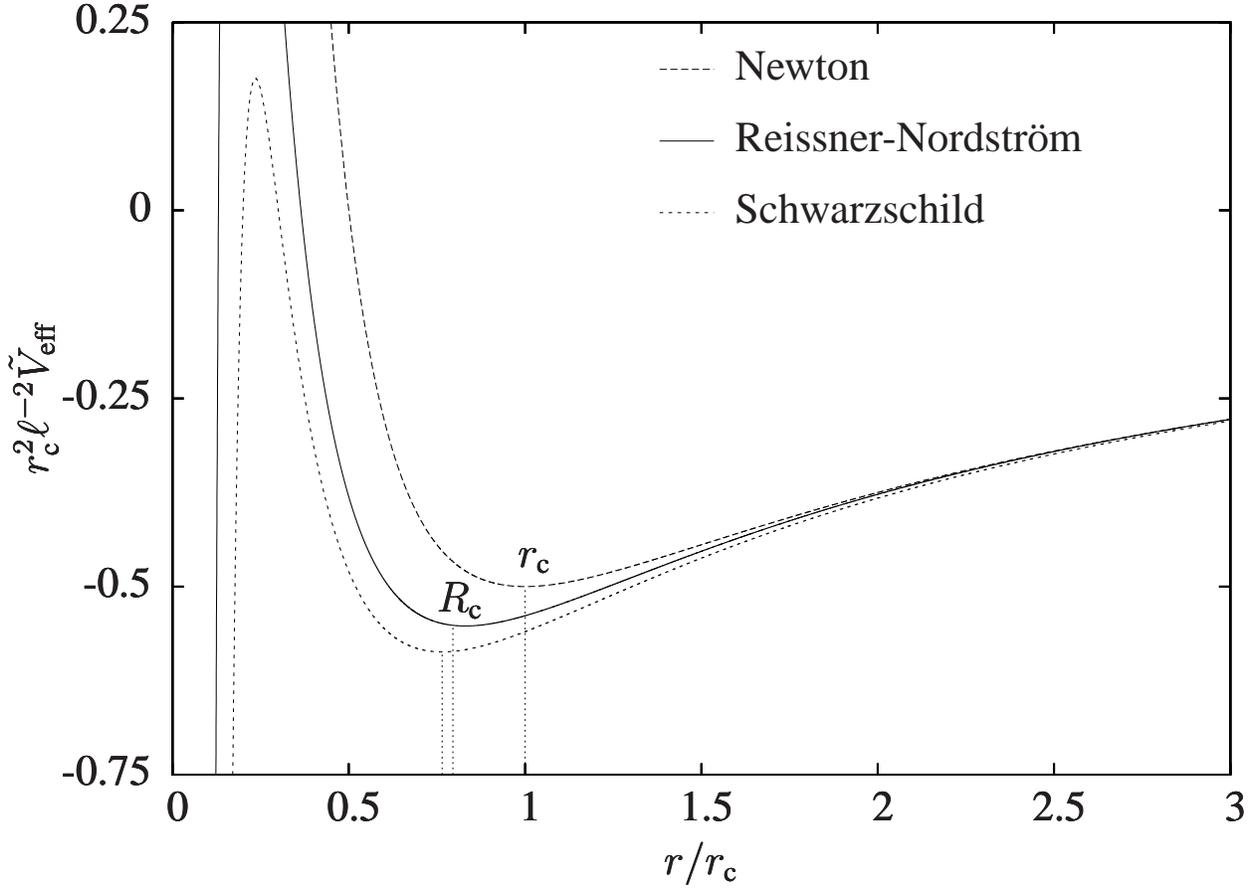}
\caption{\label{fig_RN_eff_pot} Effective potential (\ref{eq_RNEP2}) as derived from the Reissner-Nordstr\"{o}m geometry (solid) compared to that derived from Newtonian mechanics (upper, long dashes) and Schwarzschild geometry (lower, short dashes).  The vertical dotted lines identify the radius of circular orbit as predicted by (from left to right) Schwarzschild, Reissner-Nordstr\"{o}m $R_\text{c}$, and Newton $r_\text{c}$.  For the chosen values of relativistic correction parameters, the Reissner-Nordstr\"{o}m geometry predicts a radius of circular orbit (\ref{eq_RN_rc_eff_pot}) that is smaller than the corresponding Newtonian orbit and larger than the corresponding Schwarzschild orbit.  The values $\epm=0.06$ and $\epq=0.02$, for which $(\rq/\rm)^2 = 1/6$, have been chosen for purposes of illustration.  These large values are inconsistent with the condition $\epq/\epm \ll 1$ and, therefore, inconsistent with the approximation of Eq.~(\ref{eq_RN_1st_r}).  However, a self-consistent solution derived in Sec.~\ref{SubSec_RN_self_cons} correctly predicts the radius of circular Reissner-Nordstr\"{o}m orbit (\ref{eq_RN_SC_coeff_r}) for these large relativistic correction parameters.}
\end{figure}

Characteristics of Reissner-Nordstr\"{o}m orbits are described by comparison with Keplerian orbits, as in Sec.~\ref{SubSec_S_Char}.  The approximate equation of orbit (\ref{eq_RN_orbits_approx})-(\ref{eq_RN_coeff_phi}) results in the following relativistic corrections:
\begin{align}
(2\pi)^{-1}\Delta\varphi & \approx 3\epm - \epq \label{eq_RN_1st_precess};\\
\delta \tilde{r}_\text{c}/r_\text{c} &\approx - 3\epm + 2\epq \label{eq_RN_1st_r};\\
\tilde{r}_\pm/r_\pm &\approx 1 - 3\epm\frac{1 \mp 2e}{1 \mp e} + 2\epq; \\
\delta \tilde{e}/e &\approx 3\epm \label{eq_RN_1st_phi}.
\end{align}
Relativistic corrections due to charge are counter to those due to mass, except that there is no charge-related correction to eccentricity.  (The exact equation of motion (\ref{eq_RN_eoo_DE}) may also be linearized, resulting in a charge-related decrease in eccentricity of order $\epm\epq$.)  The rate of precession as predicted by the Schwarzschild geometry (\ref{eq_S_1st_precess}) is decreased when charge is present (\ref{eq_RN_1st_precess}).  This additional contribution to precession is identical to that calculated by Chaliasos \cite{chaliasos}.  (See also Teli \& Palaskar \cite{TP}.)  The radius of circular orbit as predicted by the Schwarzschild geometry (\ref{eq_S_1st_r}) is increased when charge is present (\ref{eq_RN_1st_r}).  For example, in the limit $\epq/\epm \rightarrow \tfrac{3}{2}$, $\tilde{r}_\text{c} \rightarrow r_\text{c}$.  The radius of circular orbit as determined from (\ref{eq_RN_1st_r}) is consistent to first order in $\epm$ and $\epq$ with the stable circular orbit calculated by minimizing the effective potential (\ref{eq_RNEP2}),
\begin{gather}\begin{split}
R_\text{c} ={} & \tfrac{1}{2}(1+2\epq) r_\text{c} \\
& + \tfrac{1}{2}(1+2\epq) r_\text{c} \bigl[1 - 12\epm (1 + 2 \epq)^{-2} \bigr]^\frac{1}{2}. \end{split} \label{eq_RN_rc_eff_pot}
\end{gather}
The RN effective potential (\ref{eq_RNEP2}) is compared to that derived from Newtonian mechanics (\ref{eq_N_eff_pot}) in Fig.~\ref{fig_RN_eff_pot}.  (The Schwarzschild effective potential is included for reference.)  The relativistic pericenter is reduced for $\tfrac{1}{4} < \epq/\epm < \tfrac{3}{2}$.  The remaining parameter space is considered in two cases: (Case~I) For $0 \le \epq/\epm \le \tfrac{1}{4}$, the relativistic pericenter is enlarged only for very large values of eccentricity, $\tfrac{3}{4} \le e_{-} \le 1$, for which the approximate equation of orbit (\ref{eq_RN_orbits_approx}) and resulting orbital characteristics (\ref{eq_RN_1st_precess})-(\ref{eq_RN_1st_phi}) are not expected to be accurate; (Case~II) For large values $\tfrac{3}{2} \le \epq/\epm \le 2$ it is necessary to include the higher-order $\epm\epq$-term in the equation of orbit (\ref{eq_RN_eoo_DE}) in order to make a consistent argument concerning the value of eccentricity beyond which the relativistic pericenter is  enlarged when compared to the corresponding Keplerian pericenter.  This is avoided by defining a Keplerian limit including the condition $\epq/\epm \ll 1$, or equivalently $(\rq/\rm)^2 \ll 1/2$.  (It is only necessary to impose that $\epq/\epm<\tfrac{3}{2}$.  However, it is convenient to restrict the problem to a smaller parameter space because the RN (\ref{eq_RNEP2}) and Newtonian (\ref{eq_N_eff_pot}) effective potentials intersect at $r=r_\text{i}$ given by $r_\text{c}/r_\text{i} = \epq/\epm$.  The problem is simplified if $r_\text{c}/r_\text{i} \ll 1$, in which case $r_\text{i} \gg r$ for all $r$ consistent with the domain of validity as given by (\ref{eq_linearize}).)  Having addressed both cases, it may be stated that in this Keplerian limit the relativistic pericenter is always reduced.  The relativistic apocenter is smaller than the corresponding Keplerian apocenter for eccentricities smaller than $e_+ \approx \tfrac{1}{2} - \mathcal{O} ( \epq/\epm )$.

%%%%%%%%%%%%%%%%%%%%%%%%%
\subsection{Reissner-Nordstr\"{o}m Energy Parameters}\label{SubSec_RN_energy}
%%%%%%%%%%%%%%%%%%%%%%%%%

For a particular Keplerian orbit, identified by total energy, it is necessary to identify the corresponding Reissner-Nordstr\"{o}m energy for an orbit described by the same angular momentum.  This provides a relativistic correction to Newtonian energies for Keplerian orbits.  This is also useful when comparing orbital properties using energy diagrams.  The generalized virial theorem \cite{ChaCon,bonazzola,vilain,gourgoulhon} provides this relation for circular Reissner-Nordstr\"{o}m orbits.  Referring to (\ref{eq_RN_eom}) and (\ref{eq_RN_eff_pot}), a RN potential energy parameter for a circular orbit is defined as
\begin{gather}
\tilde{V}_\text{c} \equiv -\frac{GM}{\tilde{r}_\text{c}}-\epm \frac{GMr^2_\text{c}}{\tilde{r}^3_\text{c}} + \epq \frac{GMr_\text{c}}{\tilde{r}^2_\text{c}}.
\end{gather}
According to the virial theorem,
\begin{gather}
\tilde{T}_\text{c} = \frac{GM}{2\tilde{r}_\text{c}} + 3\epm \frac{GMr^2_\text{c}}{2\tilde{r}^3_\text{c}} - \epq \frac{GMr_\text{c}}{\tilde{r}^2_\text{c}},
\end{gather}
where $\tilde{T}_\text{c}$ is a RN kinetic energy parameter.  Therefore, using $\tilde{E}_\text{c} \equiv \tilde{T}_\text{c} + \tilde{V}_\text{c}$,
\begin{align}
\tilde{E}_\text{c} &= -\frac{GM}{2\tilde{r}_\text{c}} + \epm \frac{GMr^2_\text{c}}{2\tilde{r}^3_\text{c}}\\
&= -\frac{GM}{2r_\text{c}} \frac{r_\text{c}}{\tilde{r}_\text{c}} \biggl( 1-\epm \frac{r^2_\text{c}}{\tilde{r}^2_\text{c}} \biggr) \label{eq_RN_Vir_Energy}\\
&\approx E_\text{c}( 1 + 2\epm - 2\epq ), \label{eq_RN_1st_Vir_Energy}
\end{align}
where $\tilde{r}_\text{c} \approx r_\text{c}( 1 - 3\epm + 2\epq )$ is used in the last step.  An approximate RN energy parameter is defined by ansatz as in (\ref{eq_ansatz}), resulting in a relation between noncircular RN and Keplerian energies,
\begin{gather}
\tilde{E} \approx E - [2\epm ( 1 - 4e^2 ) - 2\epq ( 1 - e^2 )]|E_\text{c}|. \label{eq_RN_1st_Energy}
\end{gather}
(This ansatz is investigated in Sec.~\ref{SubSubSec_RN_SC_energy}, wherein Eq.~(\ref{eq_RN_1st_Energy}) is shown to approximate well a more detailed parameterization and is a good approximation for near-circular orbits.)  The RN energy parameter is smaller than the corresponding Keplerian energy for eccentricities smaller than $e_0 \approx \tfrac{1}{2} + \mathcal{O} ( \epq/\epm )$.  The value of eccentricity for which the RN energy parameter is equal to the corresponding Keplerian energy is approximately the same as that for which the apocenter distances are equal, $e_0 \approx e_{+} \approx 1/2$. (See Sec.~\ref{SubSec_RN_char}.)  However, this approximate energy parametrization is not expected to be accurate for values of $e$ approaching $1/2$.  A more accurate treatment in Secs.~\ref{SubSubSec_RN_SC_char}~and~\ref{SubSubSec_RN_SC_energy} results in $e_{+} < e_0$, as expected.

%%%%%%%%%%%%%%%%%%%%%%%%%%
\subsection{Self-Consistent Keplerian Limit}\label{SubSec_RN_self_cons}
%%%%%%%%%%%%%%%%%%%%%%%%%%

A more accurate self-consistent equation of orbit is derived for the Reissner-Nordstr\"{o}m geometry by following the procedure outlined in Sec.~\ref{Sec_S_self_cons}, Eqs.~(\ref{eq_S_eoo_alt})-(\ref{eq_constant2}).  The resulting equation of motion may be expressed as
\begin{gather}
\frac{\tilde{r}_\text{c}}{r} = 1 + \tilde{e}\cos{\tilde{\kappa}\varphi}, \label{eq_RN_SC_eoo}
\end{gather}
with the following definitions:
\begin{align}
\tilde{r}_\text{c}/r_\text{c} &\equiv (1+2\epq) \biggl\{\frac{1}{2} + \frac{1}{2} \biggl[1 - \frac{12\epm}{(1 + 2 \epq)^{2}} \biggr]^\frac{1}{2} \,\biggr\}; \label{eq_RN_SC_coeff_r}\\
\tilde{e}/e &\equiv \frac{1}{1-3\epm}; \label{eq_RN_SC_coeff_e}\\
\tilde{\kappa} &\equiv (1+2\epq)^\frac{1}{2} \biggl[1 - \frac{12\epm}{(1 + 2 \epq)^{2}} \biggr]^\frac{1}{4}. \label{eq_RN_SC_coeff_phi}
\end{align}
This solution is consistent to all orders in $\epm$ and $\epq$, and the more approximate orbital equation (\ref{eq_RN_orbits_approx}) is recovered from (\ref{eq_RN_SC_eoo}) by expanding (\ref{eq_RN_SC_coeff_r})--(\ref{eq_RN_SC_coeff_phi}) and keeping terms first order in both $\epm$ and $\epq$:
\begin{align}
\tilde{r}_\text{c} &\approx r_\text{c}(1 - 3\epm + 2\epq + 6\epm\epq - 9\epmsquared);\\
\tilde{e} &\approx e(1 + 3\epm + 9\epmsquared);\\
\tilde{\kappa} &\approx 1 - 3\epm + \epq + 9\epm\epq - \tfrac{27}{2}\epmsquared - \tfrac{1}{2} \epqsquared.
\end{align}
(There is no correction of order $\epqsquared$ in the approximate radius of circular orbit $\tilde{r}_\text{c}$.)  A systematic verification may be carried out by direct substitution of Eq.~(\ref{eq_RN_SC_eoo}) into Eq.~(\ref{eq_RNeom_phi2}) using the exact definitions (\ref{eq_RN_SC_coeff_r})--(\ref{eq_RN_SC_coeff_phi}) for $\tilde{r}_\text{c}$, $\tilde{e}$, and $\tilde{\kappa}$, and discarding the term nonlinear in eccentricity.  However, the justification for discarding the term nonlinear in eccentricity is the correspondence principle.  Arguments concerning which terms to keep based only on a direct comparison of the relative magnitudes of higher-order and lower-order terms lead to contradictions.  Rather, the domain of validity is expressed by subjecting the solution (\ref{eq_RN_SC_eoo}) to condition (\ref{eq_linearize}).  Evaluating the equation of orbit (\ref{eq_RN_SC_eoo}) at pericenter results in
\begin{gather}
\frac{r_\text{c}}{\tilde{r}_{-}} = \frac{1+\tilde{e}}{\tilde{r}_\text{c}/r_\text{c}}.
\end{gather}
Therefore, according to (\ref{eq_linearize}), the domain of validity is given by
\begin{gather}
\tilde{e}/e + 2 \bigl(1 - \tilde{r}_\text{c}/r_\text{c} \bigr) \ll 1,
\end{gather}
which is consistent with Eq.~(\ref{eq_RN_1st_valid}) to first order in $\epm$ and $\epq$.  This condition reduces to that for the self-consistent Schwarzschild solution (\ref{eq_S_SC_valid}) when $\epq=0$. 

%%%%%%%%%%%%%%%%%%%%%%%%%
\subsubsection{Characteristics of Reissner-Nordstr\"{o}m Orbits}\label{SubSubSec_RN_SC_char}
%%%%%%%%%%%%%%%%%%%%%%%%%

The characteristics of Reissner-Nordstr\"{o}m orbits are expressed more accurately using the self-consistent equation of orbit (\ref{eq_RN_SC_eoo})--(\ref{eq_RN_SC_coeff_phi}).  The self-consistent equation of orbit (\ref{eq_RN_SC_eoo}) predicts the rate of precession (\ref{eq_S_def_precess}) to be
\begin{gather}\begin{split}
(2\pi)^{-1}\Delta\varphi ={} & 3\epm -\epq - 15\epm\epq \\
&+ \tfrac{45}{2}\epmsquared + \tfrac{3}{2}\epqsquared + \mathcal{O}(c^{-6}). \end{split}
\end{gather}
The effect of charge is to reduce the rate of precession predicted by the Schwarzschild geometry.

The self-consistent equation of orbit predicts the radius of stable circular orbit to be reduced by
\begin{gather}
\delta \tilde{r}_\text{c}/r_\text{c} = -3\epm + 2\epq + 6\epm\epq - 9\epmsquared + \mathcal{O}(c^{-6}).
\end{gather}
(There is no correction of order $\epqsquared$ here.)  This expression is in agreement with the Reissner-Nordstr\"{o}m geometry;  The radius of circular orbit, as predicted by this self-consistent solution (\ref{eq_RN_SC_coeff_r}), is identical to that calculated from the RN geometry (\ref{eq_RN_rc_eff_pot}) by minimizing the effective potential (\ref{eq_RNEP2}).  The relativistic apsides are
\begin{gather}
\tilde{r}_\pm/r_\pm = \frac{\tilde{r}_\text{c}}{r_\text{c}} \frac{1 \mp e}{1 \mp e/(1 - 3\epm)}.
\end{gather}
The relativistic apocenter is smaller than the corresponding Keplerian apocenter for values of eccentricity less than
\begin{gather}\begin{split}
e_{+} ={} & \tfrac{1}{2}[1 - \tfrac{1}{3}(\epq/\epm) - \tfrac{1}{9}(\epq/\epm)^2 \\
& + \tfrac{9}{2}\epmsquared - \tfrac{1}{2}\epqsquared + \mathcal{O}(\epqcubed/\epmcubed) + \mathcal{O}(c^{-6}) ]. \end{split} \label{eq_RN_SC_e_apo}
\end{gather}
(There are no corrections first order in $\epm$ and $\epq$ here.)

There is no charge-related correction to eccentricity;  $\delta\tilde{e}/e$ is identical to that derived from the Schwarzschild geometry, Eqs.~(\ref{eq_S_SC_de})~and~(\ref{eq_S_SC_de_Taylor}).

%%%%%%%%%%%%%%%%%%%%%%%%%
\subsubsection{Reissner-Nordstr\"{o}m Energy Parameters}\label{SubSubSec_RN_SC_energy}
%%%%%%%%%%%%%%%%%%%%%%%%%

A more accurate relation between the Reissner-Nordstr\"{o}m energy parameter for a circular orbit and that derived from Newtonian mechanics is constructed using the virial theorem together with the radius of circular orbit as derived from the self-consistent equation of orbit.  (See Sec.~\ref{SubSec_RN_energy}.)  The Reissner-Nordstr\"{o}m energy parameter is given in terms of the Keplerian energy by Eqs.~(\ref{eq_RN_Vir_Energy})~and~(\ref{eq_RN_SC_coeff_r}),
\begin{gather}\begin{split}
\tilde{E}_\text{c}/E_\text{c} &= \frac{(1 + 2\epq) (\tilde{r}_\text{c}/r_\text{c}) - 4\epm}{(\tilde{r}_\text{c}/r_\text{c})^3} \\
&= 1 + 2\epm - 2\epq - 12\epm\epq \\
&\quad + 9\epmsquared + 4\epqsquared + \mathcal{O}(c^{-6}), \end{split} \label{eq_RN_Vir_Energy_exact}
\end{gather}
This result is also obtained by substituting the radius of circular orbit (\ref{eq_RN_SC_coeff_r}) into the effective potential (\ref{eq_RNEP2}).  This expression is in agreement with the approximate relation (\ref{eq_RN_1st_Vir_Energy}) to first order in $\epm$ and $\epq$.

More generally, a Reissner-Nordstr\"{o}m energy parameter for noncircular, approximately Keplerian orbits is defined by
\begin{gather}
\tilde{e} \equiv \frac{R_+ - R_-}{R_+ + R_-} \approx \frac{\tilde{r}_+ - \tilde{r}_-}{\tilde{r}_+ + \tilde{r}_-},
\end{gather}
where $R_\pm=R_\pm (\tilde{E})$ are the relativistic apsides as determined from the intersection of the effective potential (\ref{eq_RNEP2}) with a line of constant energy $\tilde{E}$.  That is,
\begin{gather}
r^2_\text{c}\ell^{-2}\tilde{E} = - \frac{r_\text{c}}{R_\pm} + \frac{1}{2}\frac{r^2_\text{c}}{R_\pm^2} - \epm \frac{r^3_\text{c}}{R_\pm^3} + \epq \frac{r^2_\text{c}}{R_\pm^2}.  \label{eq_RN_cubic}
\end{gather}
The solution is then inverted to give $\tilde{E}(\tilde{e})$.  The cubic (\ref{eq_RN_cubic}) is solved using Vi\`{e}te's (1540--1603) trick \cite{fulling},
\begin{gather}
\frac{r_\text{c}}{R_\pm} = \frac{1 + 2\epq}{6\epm} \bigl[ 1 + 2 \beta \cos{( \phi_\pm + \tfrac{1}{3} \arccos{\gamma})} \bigr],
\end{gather}
where $\phi_\pm = \pi \mp \pi/3$, and the following functions are defined:
\begin{align}
\beta &\equiv \bigl[ 1 - 12\epm (1 + 2\epq)^{-2} \bigr]^\frac{1}{2};\\
\gamma &\equiv \beta^{-3} \biggl[ 1 - \frac{18\epm}{(1 + 2\epq)^{2}} \biggl( 1 - \frac{3\epm}{1 + 2\epq} \frac{\tilde{E}}{E_\text{c}} \biggr) \biggr].
\end{align}
(Notice that $2r_\text{c}^2\ell^{-2} = -1/E_\text{c}$.)  Therefore, the eccentricity is given by
\begin{gather}
\tilde{e} = \frac{\sqrt{3} \beta \sin{ (\tfrac{1}{3} \arccos{ \gamma }) }}{ 1 - \beta \cos{(\tfrac{1}{3} \arccos{\gamma})} }. \label{eq_RN_e_exact}
\end{gather}
This is verified by setting $\tilde{e}=0$ and calculating the energy of a circular RN orbit, resulting in Eq.~(\ref{eq_RN_Vir_Energy_exact}).  Inverting (\ref{eq_RN_e_exact}) results in an expression for the RN energy parameter
\begin{gather}\begin{split}
\tilde{E}/E_\text{c} ={} & \frac{1 + 2\epq}{3\epm} \biggl\{ 1 - \frac{(1 + 2\epq)^2}{18\epm} \\
& \times [ 1 - \beta^3\cos{(3\arccos{\lambda})} ] \biggr\}, \end{split} \label{eq_RN_Energy_exact}
\end{gather}
where
\begin{align}
\lambda &= \beta^{-1} \zeta^{-1} ( 1 + \sqrt{1 + \zeta\xi} ),  \\
\zeta   &= \frac{3 + \tilde{e}^2}{\tilde{e}^2}, \\
\xi	&= \frac{3\beta^2 - \tilde{e}^2}{\tilde{e}^2}.
\end{align}
Dividing (\ref{eq_RN_Energy_exact}) by (\ref{eq_RN_Vir_Energy_exact}) and expanding in powers of $\epm$ and $\epq$ results in an expression comparable to the earlier energy parameterization ansatz (\ref{eq_ansatz}),
\begin{gather}\begin{split}
\tilde{E}/\tilde{E}_\text{c} ={} &( 1 - \tilde{e}^2 ) [ 1 + 2\tilde{e}^2 \epm \\
& + \tilde{e}^2 ( 14 + 5 \tilde{e}^2 ) \epmsquared - 8\tilde{e}^2 \epm\epq ] + \mathcal{O}(c^{-6}). \end{split} \label{eq_RN_compare_to_ansatz}
\end{gather}
(There is no term of order $\epqsquared$ here.)  The Schwarzschild result (\ref{eq_S_compare_to_ansatz}) is recovered by setting $\epq = 0$, and the Keplerian result (\ref{eq_Kep_Energy}) is recovered by setting $\epm = \epq = 0$.  The RN energy parameter may instead be expressed in terms of the energy and eccentricity of the  corresponding Keplerian orbit using (\ref{eq_RN_Energy_exact}) together with (\ref{eq_RN_SC_coeff_e}) and (\ref{eq_Kep_Energy}),
\begin{gather}\begin{split}
\tilde{E}/E_\text{c} ={} & E/E_\text{c} + 2\epm ( 1 - 3e^2 - e^4 ) - 2\epq (1 - e^2) \\
& - 12\epm\epq (1 - e^2 - e^4) \\
& + \epmsquared ( 9 - 18e^2 - 37e^4 - 5e^6 ) \\
& + 4\epqsquared (1 - e^2) + \mathcal{O}(c^{-6}). \end{split} \label{eq_RN_Energy_exact_Taylor}
\end{gather}
Neglecting terms of orders $\epm e^4$\!, $\epm\epq$, $\epmsquared$, $\epqsquared$, and smaller in Eq.~(\ref{eq_RN_Energy_exact_Taylor}) results in
\begin{gather}
\tilde{E} \approx E - [2\epm ( 1 - 3e^2 ) - 2\epq ( 1 - e^2 )]|E_\text{c}|.
\end{gather}
This result is similar to the energy relation (\ref{eq_RN_1st_Energy}) derived from the simple ansatz (\ref{eq_ansatz}), lending value to the much simpler approach of Sec.~\ref{SubSec_RN_energy}.  Corrections to the orginal ansatz (\ref{eq_ansatz}) are given by Eq.~(\ref{eq_RN_compare_to_ansatz}).  The orginal ansatz (\ref{eq_ansatz}) and the resulting relation between the RN energy parameter and corresponding Keplerian energy (\ref{eq_RN_1st_Energy}) approximate well this more detailed parameterization, (\ref{eq_RN_compare_to_ansatz}) and (\ref{eq_RN_Energy_exact_Taylor}), for near-circular orbits requiring small relativistic corrections.

Reffering to (\ref{eq_RN_Energy_exact_Taylor}), the energy of a Reissner-Nordstr\"{o}m orbit is smaller than that for the corresponding Keplerian orbit, until the energies become equal for
\begin{gather}
2\epm (1 - 3e_0^2 - e_0^4) - 2\epq (1 - e_0^2) \approx 0,
\end{gather}
so that $e_0 \approx 0.55 - 0.44(\epq/\epm)$.  The relativistic energy parameter becomes greater than the corresponding Keplerian energy for $e>e_0$.  For $\epq/\epm \ll 1$ this value of eccentricity $(e_0)$ is larger than that for which the corresponding apocenter distances are equal $(e_+)$, as determined from Eq.~(\ref{eq_RN_SC_e_apo}).  For example, choosing $\epq/\epm = 1/10$ results in  $e_+ \approx 0.48$ and $e_0 \approx 0.51$.  This is expected since it is required that $\tilde{r}_{+}>r_{+}$ for any energy $\tilde{E}=E>E_\text{c}$;  Referring to an energy diagram, $\tilde{E}$ must lie below $E$ in order for $\tilde{r}_{+}=r_{+}$.

%%%%%%%%%%%%%%%%%%%%%%%%%
\section{Schwarzschild-de~Sitter Orbits in Keplerian Limit} \label{Sec_SdS_KepLim}
%%%%%%%%%%%%%%%%%%%%%%%%%

The path of a small test mass near a spherically-symmetric central mass $M$ including the effect of the cosmological constant $\Lambda$ is described by the Schwarzschild-de\,Sitter (SdS) geometry \cite{PleKra,OR,rindler,KW1,alpher,kottler, weyl,GN,SH,KHM,KW2,KKL,JS,SJ,AMF,FJ,dumin,iorio1,iorio2,HobEfsLas}.  The metric is given by Eq.~(\ref{eq_metric_general}) with
\begin{gather}
\mathrm{e}^{2\nu(r)} = \mathrm{e}^{2\mu(r)} = 1 - \rm/r - r^2/\rLsquared,
\end{gather}
where $\rLsquared \equiv 3/\Lambda$, and $\Lambda>0$ (repulsive).  A solution is described in a Keplerian limit, so that $\rL\gg r\gg \rm$.  The radius of the cosmological horizon is large when compared to the scale of the orbits.  Therefore, the singularities $\mathrm{e}^{2\mu(r^{\vphantom{0}}_{\!0})} = 0$ are irrelevant in the present context.  The equations of motion may be expressed as
\begin{align}
\ell &= r^2\dot\varphi,\\
k &= (1 - \rm/r - r^2/\rLsquared)\,\dot{t},\\
\tfrac{1}{2}(k^2-1)c^2 &=\tfrac{1}{2}\dot{r}^2 + \tilde{V}_\text{eff}, \label{eq_SdS_eom}
\end{align}
where an effective potential is defined as
\begin{gather}
r^2_\text{c}\ell^{-2}\tilde{V}_\text{eff} \!\equiv\! - \frac{r_\text{c}}{r} + \frac{1}{2}\frac{r^2_\text{c}}{r^2} - \epm \frac{r^3_\text{c}}{r^3} - \epL\frac{r^2}{r^2_\text{c}} - \frac{1}{2} \frac{r^2_\text{c}}{\rLsquared}.  \label{eq_SdS_eff_pot}
\end{gather}
The $\Lambda$-related relativistic correction parameter is defined as
\begin{gather}
\epL \equiv \frac{r_\text{c}^3}{\rm \rLsquared} = \tfrac{1}{2}\mpe \frac{r_\text{c}^2}{\rLsquared}.
\end{gather}
It is assumed that $\epL \ll \epm$, or equivalently $\Lambda \ll \rmsquared/r_\text{c}^4$.  This is reasonable considering the very small value for the cosmological constant \cite{riess1,scp,BOPS}:  $\Lambda \approx H_0^2/c^2 \approx 10^{-56}\,\text{cm}^{-2}$.  An equation for the trajectory of a test particle is obtained by eliminating time from Eq.~(\ref{eq_SdS_eom}) and differentiating once more with respect to $\varphi$,
\begin{gather}
\frac{\mathrm{d}^2}{\mathrm{d} \varphi^2} \frac{r_\text{c}}{r} + \frac{r_\text{c}}{r} = 1 + 3\epm \Big( \frac{r_\text{c}}{r} \Big)^{\!2} - 2\epL \Big( \frac{r_\text{c}}{r} \Big)^{\!-3}. \label{eq_SdS_eoo_DE}
\end{gather}
Schwarzschild orbits (\ref{eq_S_eoo_DE}) are recovered by setting $\epL=0$, and the conic-sections of Newtonian mechanics (\ref{eq_N_eoo}) are recovered by setting $\epm = \epL=0$.

Although a self-consistent solution for SdS orbits in this Keplerian limit is intractable, a first-order solution provides three orbital characteristics with much less effort.  Following the procedure used to solve Eq.~(\ref{eq_S_eoo_DE}), described by (\ref{eq_linearize})-(\ref{eq_corr_princ}), a first-order solution to (\ref{eq_SdS_eoo_DE}) may be expressed as
\begin{gather}
\frac{\tilde{r}_\text{c}}{r} \approx 1 + \tilde{e} \cos{\tilde{\kappa}\varphi}, \label{eq_SdS_eoo}
\end{gather}
where
\begin{align}
\tilde{r}_\text{c} & \approx r_\text{c} (1 - 3\epm + 2\epL), \label{eq_SdS_coeff_r} \\
\tilde{e} &\approx e (1 + 3\epm + 8\epL),\\
\tilde{\kappa} &\approx 1 - 3\epm - 3\epL. \label{eq_SdS_coeff_phi}
\end{align}
A systematic verification may be carried out by substituting (\ref{eq_SdS_eoo}) into (\ref{eq_SdS_eoo_DE}), keeping terms of orders $e$, $\epm$, $\epL$, $e\epm$, and $e\epL$ only.  However, the justification for discarding the term nonlinear in eccentricity is the correspondence principle.  Arguments concerning which terms to discard based only on direct comparisons of relative magnitudes of higher-order and lower-order terms lead to contradictions. Rather, the domain of validity is expressed by subjecting the solution (\ref{eq_SdS_eoo}) to condition (\ref{eq_linearize}) at pericenter, resulting in
\begin{gather}
e(1 + 3\epm + 8\epL) + 2(3\epm - 2\epL) \ll 1.
\end{gather}

%%%%%%%%%%%%%%%%%%%%%%%%%
\subsection{Characteristics of Schwarzschild-de~Sitter Orbits}\label{SubSec_SdS_Char}
%%%%%%%%%%%%%%%%%%%%%%%%%

Characteristics of Schwarzschild-de\,Sitter orbits are described by comparison with Keplerian orbits, as in Sec.~\ref{SubSec_S_Char}.  The approximate equation of orbit (\ref{eq_SdS_eoo})-(\ref{eq_SdS_coeff_phi}) results in the following relativistic corrections:
\begin{align}
(2\pi)^{-1}\Delta\varphi &\approx 3\epm + 3\epL \label{eq_SdS_1st_precess};\\
\delta \tilde{r}_\text{c}/r_\text{c} &\approx - 3\epm + 2\epL \label{eq_SdS_1st_r};\\
\tilde{r}_\pm/r_\pm &\approx 1-3\epm\frac{1 \mp 2e}{1 \mp e} + 2\epL \frac{1 \pm 3e}{1 \mp e}; \label{eq_SdS_GC3}\\
\delta \tilde{e}/e &\approx 3\epm + 8\epL \label{eq_SdS_1st_phi}.
\end{align}
Relativistic corrections due to $\Lambda$ include both increased rate of precession and increased eccentricity.  (Recall from Sec.~\ref{SubSec_RN_char} that there is no correction to eccentricity due to charge.)  The rate of precession as predicted by the Schwarzschild geometry (\ref{eq_S_1st_precess}) is increased if $\Lambda$ is present (\ref{eq_SdS_1st_precess}).  The additional contribution, $(2\pi)^{-1}\Delta\varphi^{}_\Lambda \approx 3\epL$, is in agreement with the standard perturbative result \cite{alpher,rindler}.  The eccentricity as predicted by the Schwarzschild geometry (\ref{eq_S_de}) is increased if $\Lambda$ is present (\ref{eq_SdS_1st_phi}).  Another effect of $\Lambda$ is to increase the overall size of the orbit, when compared to Schwarzschild orbits.  The radius of circular orbit as predicted by the Schwarzschild geometry (\ref{eq_S_1st_r}) is increased when $\Lambda$ is present (\ref{eq_SdS_1st_r}).  The present formalism provides a determination of the radius of relativistic circular orbit to first order in $\epm$ and $\epL$ (\ref{eq_SdS_coeff_r});  the standard approach of minimizing the effective potential results in a quintic equation for $\tilde{r}_\text{c}$.

The relativistic pericenter (\ref{eq_SdS_GC3}) is always reduced.  The condition for which $\tilde{r}_- = r_-$ is \mbox{$\epL/\epm \ge 3/2$,} which is inconsistent with the stated assumption of this Keplerian limit: $\epL/\epm \ll 1$.  When compared to a Schwarzschild orbit the effect of $\Lambda$ is to increase the pericenter for values $e<1/3$ and to decrease the pericenter for values $e>1/3$.  The relativistic apocenter is smaller than the corresponding Keplerian apocenter for eccentricities smaller than $e_+ \approx \tfrac{1}{2} - \mathcal{O} ( \epL/\epm )$.

%%%%%%%%%%%%%%%%%%%%%%%%%
\subsection{Schwarzschild-de~Sitter Energy Parameters}\label{SubSec_SdS_Energy}
%%%%%%%%%%%%%%%%%%%%%%%%%

For a particular Keplerian orbit, identified by total energy, it is necessary to identify the corresponding Schwarzschild-de\,Sitter energy for an orbit described by the same angular momentum.  This provides a relativistic correction to Newtonian energies for Keplerian orbits.  This is also useful when comparing orbital properties using energy diagrams.  The generalized virial theorem \cite{ChaCon,bonazzola,vilain,gourgoulhon} provides this relation for circular Schwarzschild-de\,Sitter orbits.  Referring to (\ref{eq_SdS_eom}) and (\ref{eq_SdS_eff_pot}), a SdS potential energy parameter for a circular orbit is defined as
\begin{gather}
\tilde{V}_\text{c} \equiv -\frac{GM}{\tilde{r}_\text{c}}-\epm \frac{GMr^2_\text{c}}{\tilde{r}^3_\text{c}} - \epL \frac{GM \tilde{r}^2_\text{c}}{r^3_\text{c}}.
\end{gather}
According to the virial theorem,
\begin{gather}
\tilde{T}_\text{c} = \frac{GM}{2\tilde{r}_\text{c}} + 3\epm \frac{GMr^2_\text{c}}{2\tilde{r}^3_\text{c}} - \epL \frac{GM \tilde{r}^2_\text{c}}{r^3_\text{c}},
\end{gather}
where $\tilde{T}_\text{c}$ is a SdS kinetic energy parameter.  Therefore, using $\tilde{E}_\text{c} \equiv \tilde{T}_\text{c} + \tilde{V}_\text{c}$,
\begin{align}
\tilde{E}_\text{c} &= -\frac{GM}{2\tilde{r}_\text{c}} + \epm \frac{GMr^2_\text{c}}{2\tilde{r}^3_\text{c}} - 2\epL \frac{GM \tilde{r}^2_\text{c}}{r^3_\text{c}}\\
&= -\frac{GM}{2r_\text{c}} \frac{r_\text{c}}{\tilde{r}_\text{c}} \Bigl( 1-\epm \frac{r^2_\text{c}}{\tilde{r}^2_\text{c}} + 4\epL \frac{\tilde{r}^3_\text{c}}{r^3_\text{c}} \Bigr)\\
&\approx E_\text{c}( 1 + 2\epm + 2\epL ),
\end{align}
where $\tilde{r}_\text{c} \approx r_\text{c}( 1 - 3\epm + 2\epL )$ is used in the last step.  An approximate SdS energy parameter is defined by ansatz as in (\ref{eq_ansatz}), resulting in a relation between noncircular SdS and Keplerian energies,
\begin{gather}
\tilde{E} \approx E - [2\epm ( 1 - 4e^2 ) + 2\epL ( 1 - 9e^2 )]|E_\text{c}|.
\end{gather}
The SdS energy parameter is smaller than the corresponding Keplerian energy for eccentricities smaller than $e_0 \approx \tfrac{1}{2} - \mathcal{O} ( \epL/\epm )$.  The value of eccentricity for which the SdS energy parameter is equal to the corresponding Keplerian energy is approximately the same as that for which the apocenter distances are equal, $e_0 \approx e_{+} \approx 1/2$. (See Sec.~\ref{SubSec_SdS_Char}.)  However, this approximate energy parametrization is not expected to be accurate for values of $e$ approaching $1/2$.

%%%%%%%%%%%%%%%%%%%%%%%%%
\section{Summary} \label{Sec_Summary}
%%%%%%%%%%%%%%%%%%%%%%%%%

The relativistic central-mass problem is investigated in a Keplerian limit.  Beginning with the Schwarzschild metric, a relativistic equation of orbit is derived that is similar in form to that describing Keplerian orbits.  This equation of orbit includes three relativistic corrections to Keplerian orbits: precession; reduced radial coordinate; and increased eccentricity.  The prediction for the relativistic contribution to precession is in agreement with existing calculations and observations.  The predicted reduction in size of a circular orbit is also in agreement with existing calculations.  These agreements provide confidence in a new quantitative prediction of increased eccentricity,  which may be subjected to observational tests.

The methods and approximations describing this Keplerian limit to the Schwarzschild geometry are also applied to the Reissner-Nordstr\"{o}m and Schwarzschild-de\,Sitter metrics.  The resulting equations of orbit are identical in form to that derived for the Schwarzschild metric and include relativistic corrections due to charge (RN) and cosmological constant (SdS).  In every case the relativistic equation of orbit has a form that is easily compared to that describing Keplerian orbits (\ref{eq_N_eoo}) of Newtonian mechanics,
\begin{gather}
\frac{\tilde{r}_\text{c}}{r} = 1 + \tilde{e} \cos{\tilde{\kappa}\varphi}. \label{eq_summary1}
\end{gather}
The coefficients $\tilde{r}_\text{c}$, $\tilde{e}$, and $\tilde{\kappa}$ provide relativistic corrections for each geometry.  These corrections are given to first order in Table~\ref{tab_summary1}.
\begin{table}[ht] \label{tab_summary1}
\caption{Summary of first-order relativistic corrections to Keplerian (K) orbits for the Schwarzschild (S), Reissner-Nordstr\"{o}m (RN), and Schwarzschild-de\,Sitter (SdS) geometries.  The coefficients of the relativistic equation of orbit, Eq.~(\ref{eq_summary1}), are given for each of the three geometries.  In the bottom two rows the subscript $X$ represents mass $M$, charge $Q$, or cosmological constant $\Lambda$.  The parameter $r_\text{c} \equiv \ell^2/GM$ is the radius of a circular Keplerian orbit with angular momentum $\ell$.}
\vspace*{-4ex}
\begin{equation*}
\begin{array}{c||c|c|c|c}
\hline\hline
{} & \text{\,K\,} & \text{S} & \text{RN} & \text{SdS} \\ \hline\hline
\tilde{\kappa} \vphantom{\Bigr]} & 1 & 1 - 3\epm & 1 - 3\epm + \epq & 1 - 3\epm - 3\epL \\ \hline
\tilde{r}_\text{c}/r_\text{c} \vphantom{\Bigr]} & 1 & 1 - 3\epm & 1 - 3\epm + 2\epq & 1 - 3\epm + 2\epL \\ \hline
\tilde{e}/e \vphantom{\Bigr]} & 1 & 1 + 3\epm & 1 + 3\epm & 1 + 3\epm + 8\epL \\ \hline
\epsilon^{}_{\!X} \vphantom{\Biggr]^2_2} & \text{--} & \epm \equiv \dfrac{\rm}{2 r_\text{c}}  & \epq \equiv \dfrac{\rqsquared}{\rm r_\text{c}} & \epL \equiv \dfrac{r_\text{c}^3}{\rm \rLsquared} \\ \hline
r^{}_{\!X} \vphantom{\Biggr]^2_2} & \text{--} & \rm \equiv \dfrac{2GM}{c^2} & \rq \equiv \Bigl( \dfrac{GQ^2}{4\pi\varepsilon_0 c^4}\Bigr)^{\!1/2} & \rL \equiv \Bigl( \dfrac{3}{\Lambda}\Bigr)^{\!1/2} \\ \hline\hline
\end{array}
\end{equation*}
\end{table}
\noindent The first-order shift in apside (precession) per revolution $\Delta\varphi = 2\pi (\tilde{\kappa}^{-1} -1)$ is consistent with known results for the Schwarzschild,  Reissner-Nordstr\"{o}m, and Schwarzschild-de\,Sitter geometries.  (See Table~\ref{tab_summary2}, left column.)  The RN geometry predicts an additional contribution to precession in the opposite direction to that due to matter, while the SdS geometry predicts an additional contribution to precession in the same direction as that due to matter.   For each of the three geometries, the radius of circular orbit is consistent to first order with that determined by minimizing the relativistic effective potential.  The Schwarzschild geometry predicts a reduced radius of circular orbit.  Both the RN and SdS geometries predict an radius of circular orbit that is larger than that predicted by the Schwarzschild geometry, but still smaller than that for a corresponding Keplerian orbit.  Schwarzschild orbits are predicted to be more eccentric than corresponding Keplerian orbits, and a cosmological constant (SdS) serves to further increase the eccentricity.  The presence of electric charge (RN) does not result in any additional contribution to eccentricity.  Relativistic corrections to eccentricity may serve as additional tests of general relativity.
\begin{table}[ht]\label{tab_summary2}
\caption{Summary of additional first-order relativistic corrections to Keplerian (K) orbits for the Schwarzschild (S), Reissner-Nordstr\"{o}m (RN), and Schwarzschild-de\,Sitter (SdS) geometries.  First-order angular shifts in apsides (precession) per revolution are listed in the left column.  First-order relativistic corrections to apsides are listed in the right column.  The relativistic correction parameters $\epsilon^{}_{\!X}$ are listed in Table~\ref{tab_summary1}, bottom two rows.}
\vspace*{-4ex}
\begin{equation*}
\begin{array}{c||c|c}
\hline\hline
\vphantom{\biggr]} & (2\pi)^{-1}\Delta\varphi & \tilde{r}_\pm/r_\pm \\ \hline\hline
\text{K} & 0 & 1 \\ \hline
\text{S} \vphantom{\biggr]^2_2} & 3\epm & 1 - 3\epm\dfrac{1 \mp 2e}{1 \mp e} \\ \hline
\text{RN} \vphantom{\biggr]^2_2} & 3\epm - \epq & 1 - 3\epm\dfrac{1 \mp 2e}{1 \mp e} + 2\epq \\ \hline
\text{SdS} \vphantom{\biggr]^2_2} & 3\epm + 3\epL & 1 - 3\epm\dfrac{1 \mp 2e}{1 \mp e} + 2\epL \dfrac{1 \pm 3e}{1 \mp e} \\
\hline\hline
\end{array}
\end{equation*}
\end{table}

This model and the resulting first-order corrections are valid for near-circular orbits $(e \ll 1/2)$ that require only small relativistic corrections $(\epm \ll 1/12; \epq \ll \epm; \epL \ll \epm)$.  In addition to the properties listed in Table~\ref{tab_summary1},  the overall size of a Schwarzschild orbit is predicted to be smaller than a corresponding Keplerian orbit.  This is determined not only by the radius of circular orbit, but also by a comparision of relativistic apsides to those for corresponding Keplerian orbits.  Both the apocenter and pericenter distances are found to be smaller for Schwarzschild orbits.  (See Table~\ref{tab_summary2}, right column.)  Both the RN and SdS geometries predict apsides that are larger than the Schwarzschild apsides, but still smaller than the corresponding Keplerian apsides.

\begin{figure}
\centering
\includegraphics[width=\columnwidth]{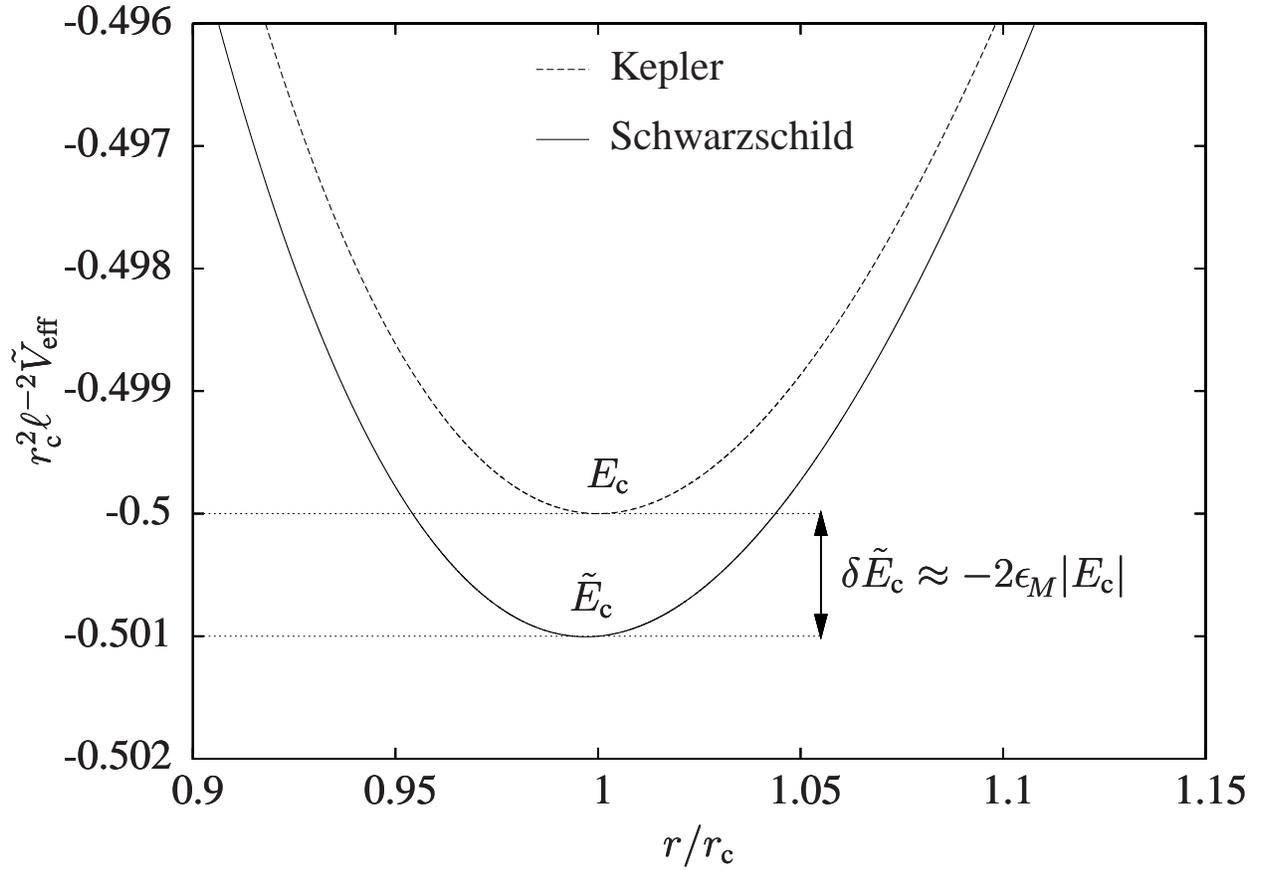}
\caption{\label{fig_dE_circular} Effective potential (\ref{eq_S_eff_pot}) derived from the Schwarzschild geometry (solid) compared to that derived from Newtonian mechanics (dashed).  Orbital energies are superimposed using dotted horizontal lines; The energy for a circular orbit as predicted by Newtonian mechanics (top) is greater than the Schwarzschild energy parameter (bottom) describing the corresponding relativistic circular orbit.  The energies are related approximately by Eq.~(\ref{eq_S_1st_Energy_circ}).  The value $\epm=10^{-3}$ has been chosen.}
\end{figure}

\begin{figure}
\centering
\includegraphics[width=\columnwidth]{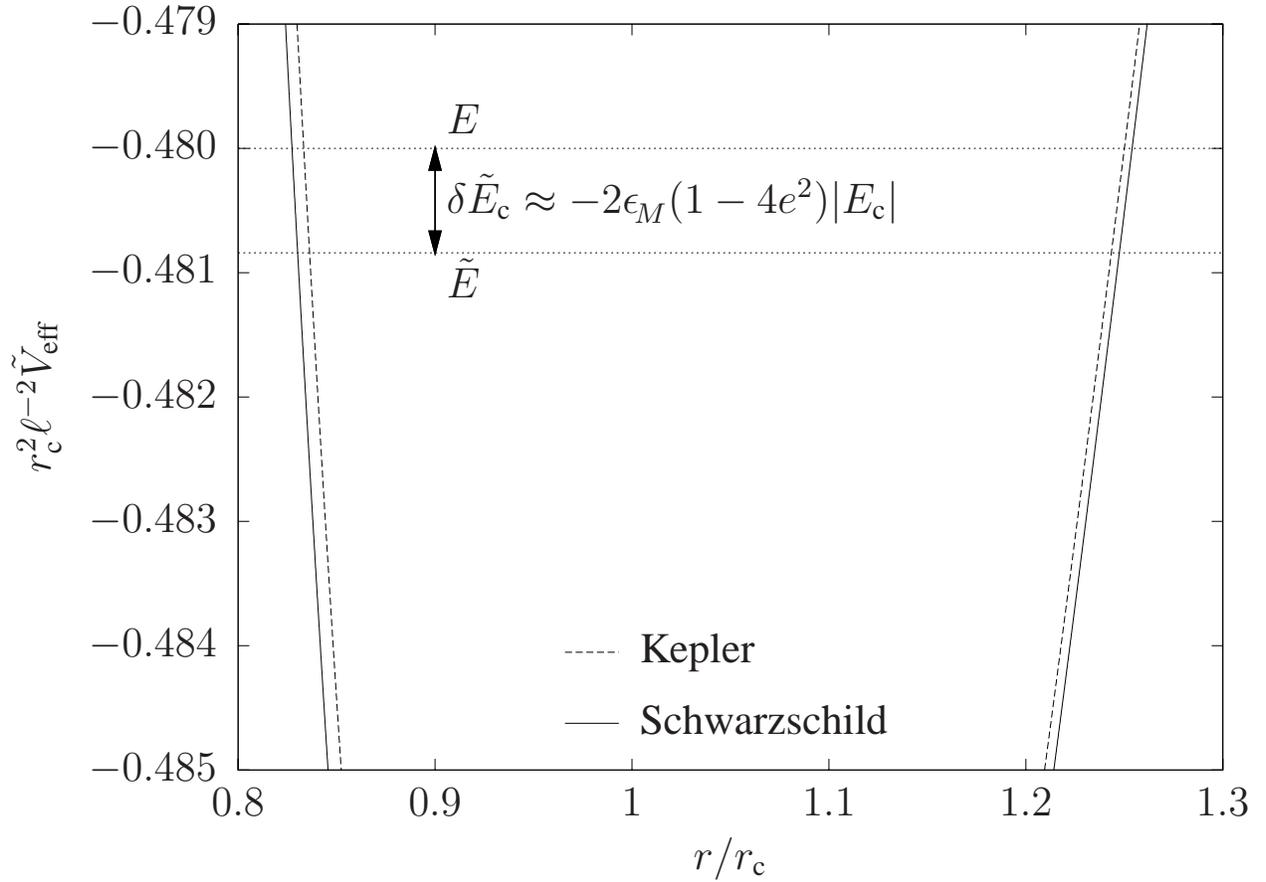}
\caption{\label{fig_dE} Effective potential (\ref{eq_S_eff_pot}) derived from the Schwarzschild geometry (solid) compared to that derived from Newtonian mechanics (dashed).  Orbital energies are superimposed using dotted horizontal lines; For the chosen value of eccentricity, the energy for an elliptical orbit as predicted by Newtonian mechanics (top) is greater than the Schwarzschild energy parameter (bottom) describing the corresponding noncircular Schwarzschild orbit.  The energies are related approximately by Eq.~(\ref{eq_S_1st_Energy}).  The values $e=0.2$ and \hbox{$\epm=10^{-3}$} have been chosen.}
\end{figure}

\begin{figure}
\centering
\includegraphics[width=\columnwidth]{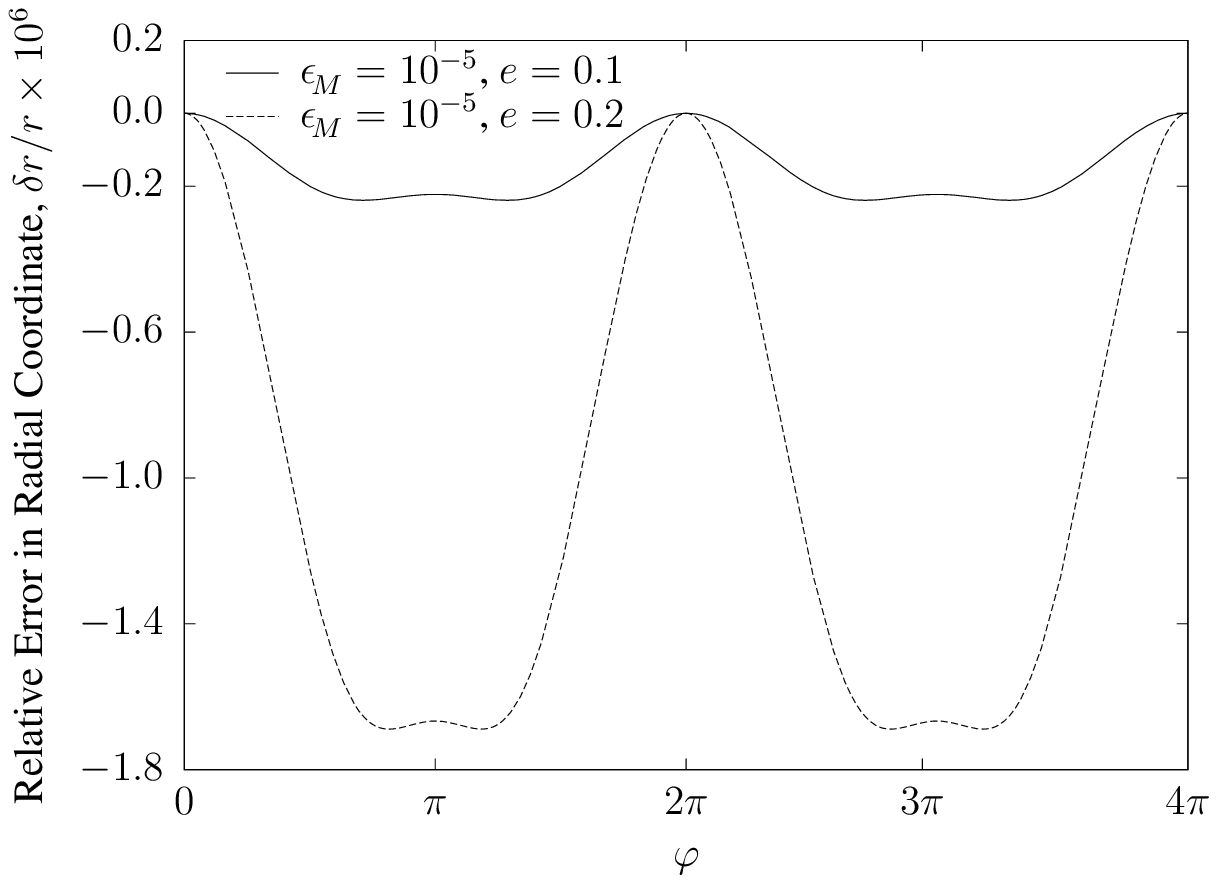}
\caption{\label{fig_moderate} Comparison of self-consistent Schwarzschild solution (\ref{eq_S_SC_eoo}) to the exact numerical solution of (\ref{eq_S_eoo_DE}) for $\epm=10^{-5}$ and moderate values of $e$.  The relative error, Eq.~(\ref{eq_rel_err_r}), is plotted for the initial interval $0 \le\varphi\le 4\pi$.  The phases are nearly equal over this initial small interval, so that this figure effectively represents the relative error in the radial coordinate:  $\delta r/r \sim 10^{-7}$ for $e=0.1$; and $\delta r/r \sim 10^{-6}$ for $e=0.2$.  In each of these two cases the relative error in angular frequency, Eq.~(\ref{eq_rel_err_phi}), is found to be: $\delta \kappa/\kappa \sim 10^{-10}$.}
\end{figure}

Long-term orbital behavior is predicted very accurately using a self-consistent Keplerian limit.  This is demonstrated by comparing a self-consistent equation of orbit to the exact numerical solution for Schwarzschild orbits.  The self-consistent equation of orbit is also given by Eq.~(\ref{eq_summary1}), but with more accurate expressions for the coefficients $\tilde{r}_\text{c}$, $\tilde{e}$, and $\tilde{\kappa}$.  For examples, the self-consistent solution predicts: a radius of circular orbit that is identical to that calculated by minimizing the relativistic effective potential; and a relative error in angular frequency of $~10^{-10}$ over 1600\,cycles.  (See Fig.~\ref{fig_moderate}.)  This solution is accurate in describing Schwarzschild orbits with eccentricities as large as $e=1/2$ and requiring large relativistic corrections.  The coefficients in Table~\ref{tab_summary1} are found to be limiting cases of those derived for the more accurate self-consistent equation of orbit, lending value to the simpler approach.  A self-consistent equation of orbit is also derived for the Reissner-Nordstr\"{o}m geometry, foregoing numerical studies.  A self-consistent model for the Schwarzschild-de\,Sitter geometry is intractable and is not pursued.
\begin{table}[ht]\label{tab_summary3}
\caption{Summary of relations between Newtonian energies for Keplerian (K) orbits and Schwarzschild (S), Reissner-Nordstr\"{o}m (RN), and Schwarzschild-de\,Sitter (SdS) energy parameters.  First-order relativistic corrections to energies for circular orbits are listed in the left column.  First-order relativistic corrections to energies for near-Keplerian orbits are listed in the right column.  The relativistic correction parameters $\epsilon^{}_{\!X}$ are listed in Table~\ref{tab_summary1}, bottom two rows.}
\vspace*{-4ex}
\begin{equation*}
\begin{array}{c||c|c}
\hline\hline
\vphantom{\biggr]} & \tilde{E}_\text{c}/E_\text{c} & (\tilde{E} - E)/|E_\text{c}| \\ \hline\hline
\text{K} & 1 & 0 \\ \hline
\text{S} \vphantom{\Bigr]} & 1 + 2\epm & -2\epm (1 - 4e^2) \\ \hline
\text{RN} \vphantom{\Bigr]} & 1 + 2\epm - 2\epq & -2\epm (1 - 4e^2) + 2\epq (1 - e^2) \\ \hline
\text{SdS} \vphantom{\Bigr]} & 1 + 2\epm + 2\epL & -2\epm (1 - 4e^2) - 2\epL (1 - 9e^2) \\
\hline\hline
\end{array}
\end{equation*}
\end{table}
Very accurate relativistic energy parameters for circular orbits are derived using the virial theorem.  This is useful for comparing the energy of a circular relativistic orbit $\tilde{E}_\text{c}$ to the energy of a corresponding circular Keplerian orbit $E_\text{c}$.  This relation is summarized for the three geometries in Table~\ref{tab_summary3}, left column.  (See also Fig.~\ref{fig_dE_circular}.)  Because the relativistic orbits are taken to be very near-Keplerian, an energy parameterization for noncircular bound orbits analogous to that for Newtonian mechanics is investigated.  The total energy of a relativistic orbit $\tilde{E}$ is defined by simple ansatz,
\begin{gather}
\tilde{E} = (1 - \tilde{e}^2)\tilde{E}_\text{c}, \label{eq_summary2}
\end{gather}
where $\tilde{E}_\text{c}$ is the energy of a relativistic circular orbit, and $\tilde{e}$ is a new relativistic eccentricity derived in the context of the relativistic equation of orbit (\ref{eq_summary1}).  Then, using the relation between $\tilde{e}$ and $e$ (Table~\ref{tab_summary1}, middle row), together with the relation between $\tilde{E}_\text{c}$ and $E_\text{c}$ (Table~\ref{tab_summary3}, left column), a relation between the total energy for a noncircular relativistic orbit and total energy for a Keplerian orbit is derived.  This relation is summarized for the three geometries in Table~\ref{tab_summary3}, right column.  (See also Fig.~\ref{fig_dE}.)  The virial theroem and simple ansatz (\ref{eq_summary2}) provide first-order relativistic corrections to Newtonian energies for bound orbits.  Finally, this simple energy parameterization is compared to a more detailed parameterization constructed using the intersection of the relativistic effective potential with a line of constant energy.  The results are similar, lending value to the simple ansatz (\ref{eq_summary2}) and resulting approximate relations.  A more detailed energy parameterization for the Schwarzschild-de\,Sitter geometry results in a quintic equation for $\tilde{E}$ and is not pursued.

The additional unstable circular orbit and relativistic capture described in the standard Newtonian limit to general relativity \cite{MTW,wald,hartle,carroll2,rindler,HobEfsLas,OR} are absent in the present treatment.  However, the present approach to the relativistic central-mass problem results in an equation of orbit that exhibits several characteristics of relativistic orbits at once.  In this Keplerian limit characteristics of general-relativistic orbits are provided as corrections to Keplerian orbits of Newtonian mechanics, providing a qualitative and quantitative understanding of the effects of general relativity on bound systems.  It should also be possible to adapt these results to relative motion of binary systems \cite{JS,iorio1}.  Perhaps more general statements may be made concerning larger systems and more extreme environments.  Globular clusters are expected to be biased toward high eccentricity in galaxies with larger cores.  Individual stars orbiting near blackholes are expected to be in anomolously small and eccentric orbits.  These characteristics could also serve as an indicator of dark matter and dark energy.  Consider two galaxies, each having approximately the same amount of visible matter.  A large difference in the amount of coexisting dark matter should be apparent in the eccentricities of the orbits of individual stars and star clusters.  Individual outlying members of galaxy clusters are expected to be biased toward high eccentricity and large precession rates due to both the large central mass and the cosmological constant.  Although effects of the cosmological constant on planetary orbits have been ruled out, they may be observable in galaxy clusters with larger radii, for which $\epL \sim \Lambda r_\text{c}^2$ becomes non-negligible.

The methods and approximations describing this Keplerian limit may be applied to other static spherical spacetimes.  The results summarized in Tables~\ref{tab_summary1}, \ref{tab_summary2} and \ref{tab_summary3} already describe related geometries using simple replacements.  The results for the Reissner-Nordstr\"{o}m geometry are extended to include magnetic charge with the replacement $Q^2/\varepsilon_0 \rightarrow Q^2/\varepsilon_0 + P^2\mu_0 c^2$, where $P$ is the magnetic charge \cite{carroll2,PleKra}.  The results for the Schwarzschild-de\,Sitter geometry become those for the Schwarzschild-anti-de\,Sitter $(\Lambda < 0)$ with the replacement $\epL \rightarrow -\epL$ \cite{KW2,SH,SH2,COV}.  It may also be possible to apply this Keplerian limit to more exotic objects such as wormholes \cite{SR}, naked singularities, and Boson and Fermion stars \cite{perlick}.

%%%%%%%%%%%%%%%%%%%%%%%%%
\begin{acknowledgments}
The authors would like to thank Neil Ashby, Shane Burns, Kristine Lang, Katherine Mondragon, Patricia Purdue, and Mauri Valtonen for their valuable comments, suggestions, and corrections.
\end{acknowledgments}
%%%%%%%%%%%%%%%%%%%%%%%%%

\pagebreak

%%%%%%%%%%%%%%%%%%%%%%%%%
%%%%%%%%%%%%%%%%%%%%%%%%%

%%%%%%%%%%%%%%%%%%%%%%%%%
%%%%%%%%%%%%%%%%%%%%%%%%%

%%%%%%%%%%%%%%%%%%%%%%%%%
%%%%%%%%%%%%%%%%%%%%%%%%%

%%%%%%%%%%%%%%%%%%%%%%%%%
%%%%%%%%%%%%%%%%%%%%%%%%%

%%%%%%%%%%%%%%%%%%%%%%%%%
%%%%%%%%%%%%%%%%%%%%%%%%%

%%%%%%%%%%%%%%%%%%%%%%%%%
%%%%%%%%%%%%%%%%%%%%%%%%%

%%%%%%%%%%%%%%%%%%%%%%%%%
%%%%%%%%%%%%%%%%%%%%%%%%%

%%%%%%%%%%%%%%%%%%%%%%%%%
%%%%%%%%%%%%%%%%%%%%%%%%%

%%%%%%%%%%%%%%%%%%%%%%%%%
%%%%%%%%%%%%%%%%%%%%%%%%%
\end{document}